\documentclass[12pt]{article}
\usepackage{amsmath,amssymb,amsfonts,color,graphicx,cite}
\usepackage{latexsym}
\usepackage{orcidlink}
\usepackage[normalem]{ulem}
\usepackage{epsfig,psfrag,rotating,soul}
\usepackage{rotfloat}
\def\reffi#1{\mbox{Fig.~\ref{#1}}}
\def\refeq#1{\mbox{Eq.~(\ref{#1})}}

\def\citere#1{\mbox{Ref.~\cite{#1}}}
\def\citeres#1{\mbox{Refs.~\cite{#1}}}

\newcommand{\tev}{\,\, \mathrm{TeV}}
\newcommand{\gev}{\,\, \mathrm{GeV}}

\def\refeq#1{\mbox{Eq.~(\ref{#1})}}

\def\reffi#1{\mbox{Fig.~\ref{#1}}}

\def\refta#1{\mbox{Tab.~\ref{#1}}}
\def\refse#1{\mbox{Sect.~\ref{#1}}}

\def\citere#1{\mbox{Ref.~\cite{#1}}}
\def\citeres#1{\mbox{Refs.~\cite{#1}}}

\evensidemargin \oddsidemargin
\allowdisplaybreaks

\begin{document}
\thispagestyle{empty}

\def\thefootnote{\fnsymbol{footnote}}


\vspace{0.5cm}

\begin{center}

\begin{large}
\textbf{The Muon Anomalous Magnetic Moment in the}
\\[2ex]
\textbf{Excited Fermion Paradigm}
\end{large}

\vspace{1cm}

{\sc
M.~Rehman\orcidlink{0000-0002-1069-0637},$^{1}$%
\footnote{email: m.rehman@comsats.edu.pk}%
,~H.~Muhammad$^{1}$%
\footnote{email:hajicomsats7515@gmail.com}%
,~O.~Panella\orcidlink{0000-0003-4262-894X},$^{2}$%
\footnote{email:orlando.panella@cern.ch}%
~and M.E.~G{\'o}mez\orcidlink{0000-0002-0137-0295}$^{3}$%
\footnote{email: mario.gomez@dfa.uhu.es}%
}

\vspace*{.7cm}
{\sl
$^1$Department of Physics, Comsats University Islamabad, 44000
  Islamabad, Pakistan \\[.1em]
$^2$ INFN, Sezione di Perugia, Via A. Pascoli, I-06123, Perugia, Italy \\[.1em]
$^3$Dpt. de Ciencias Integradas y Centro de Estudios Avanzados en F\'{i}sica Matem\'aticas y Computaci\'on, Campus del Carmen, Universidad de Huelva, Huelva 21071, Spain \\

}
\end{center}

\vspace*{0.1cm}

\begin{abstract}
\noindent

Extensions of the Standard Model featuring excited fermions present an interesting framework that motivates the search for exotic particles at the LHC. Additionally, these extensions offer potential explanations for the muon's anomalous magnetic moment and other precision observables, shedding light on the energy scale and key parameters of the new theory. Our analysis focuses on the one-loop radiative correction originating from excited lepton doublet and triplet states using  the effective Lagrangian approach. The bounds derived from the $(g-2)_\mu$ anomaly can be complemented with the ones arising from other observables like  the electroweak precision observable $\Delta \rho$ and the signals from  direct LHC searches to constraint the effective theory. Our results suggest that the $(g-2)_\mu$ anomaly can be addressed within a very narrow region of the effective theory scale. Consequently, this imposes indirect constraints on the parameter space of excited fermions.

\end{abstract}

\def\thefootnote{\arabic{footnote}}
\setcounter{page}{0}
\setcounter{footnote}{0}

\newpage


\section{Introduction}
\label{sec:level1}

The experimental measurements of the muon's magnetic moment $(g-2)_{\mu}$ have shown a persistent and statistically significant deviation from the standard model (SM) predicted value. The latest evaluation by the Muon $g-2$ collaboration at Fermilab \cite{Muong-2:2023cdq, Muong-2:2021ojo}, when combined with earlier results from the Brookhaven E821 experiment \cite{Muong-2:2006rrc}, indicates a deviation of $5.1\sigma$ from the SM prediction \cite{Aoyama:2020ynm}. The $(g-2)_\mu$ anomaly has generated considerable interest in the scientific community as it may potentially lead to the presence of new physics beyond the SM\cite{Athron:2021iuf}. 

Apart from the $(g-2)_\mu$ anomaly, there are lingering questions that remain unanswered within the framework of the SM, despite its notable agreement with experimental data. Notably, the SM falls short in explaining the existence of three generations of fermions and the observed patterns in fermion masses. Answering such questions becomes feasible under the assumption of the composite structure of fermions. This presupposition entails that the SM is a limiting case of a more fundamental theory, valid up to a certain high-energy scale denoted as the composite scale $\Lambda$. The concept of compositeness predicts the existence of heavy excited particles, each corresponding to a fermion state with a mass denoted as $M$. 

Numerous endeavors have been undertaken to explore physics at the composite scale, with a predominant focus on the production of excited fermions at colliders. Notably, studies ~\cite{Baur:1989kv,Baur:1987ga} concentrated on the production of excited states belonging to multiplets with isospin $I_W=0, 1/2$. Bounds on the masses of excited fermions with $I_W=0,1/2$ were presented in~\cite{ATLAS:2010ivc, ATLAS:2015uhg} based on experimental searches at the LHC. On the phenomenological front, calculations of excited fermion contributions to $Z$ pole observables were conducted in~\cite{Gonzalez-Garcia:1996oxz} for isospin doublet states. It has been demonstrated that higher isospin multiplets up to $I_W=1,3/2$ are permitted by SM symmetries~\cite{Pancheri:1984sm}, implying the potential existence of exotic states such as quarks $U^+$ with a charge of $+5/3e$ and quarks $D^-$ with a charge of $-4/3 e$. Phenomenological studies exploring these exotic states have been presented in ~\citeres{Biondini:2012ny,Leonardi:2014epa,Biondini:2014dfa,Leonardi:2015qna}.

Experimental searches have imposed stringent constraints on the composite scale $\Lambda$ and masses $M$ of excited fermions\cite{CMS:2018sfq, CMS:2022chw, ATLAS:2023kek}, yet direct experimental confirmation of the existence of such states remains elusive. In the absence of direct detection, the examination of indirect effects of excited fermions on SM observables proves to be a valuable probe for understanding these states. Lately, there have been intriguing advancements in exploring the phenomenology of effective interactions of excited fermions, achieved through the computation of unitarity bounds~\cite{Biondini:2019tcc}. These unitarity bounds have demonstrated significant potential when contrasted with constraints derived from direct searches at colliders~\cite{Biondini:2019tcc}. Likewise, in ~\cite{Rehman:2020ana}, it was demonstrated that non-universal contributions to electroweak precision observables $\Delta\rho$ could considerably constrain the parameter space, particularly if the masses of the excited fermions exhibit non-degeneracy.

The effects of excited fermions with isospin doublets $I_W=1/2$ on the muon's magnetic moment $(g-2)_{\mu}$ were discussed in ~\cite{Brodsky:1980zm,Renard:1982ij,Terazawa:1976xx, Terazawa:1980nck,  Gonzalez-Garcia:1996wkh}. In this paper, we extend the previous work on the subject and explore effects of excited fermions with isospin doublets $I_W=1/2$ as well as  isospin triplets $I_W=1$ on the muon's magnetic moment $(g-2)_{\mu}$ at the one-loop level. The couplings of the excited fermions (leptons) of the triplet to both excited and standard fermions (leptons) were calculated using an effective field theory approach in our previous work~\cite{Rehman:2020ana}, which will be detailed in \refse{model_setup} along with a brief description of the excited fermion model. The analytical outcomes for the contribution of excited fermions (leptons) to $(g-2)_{\mu}$ will be presented in ~\refse{Aresults}. Our numerical analysis will be provided in ~\refse{Nresults}, and our conclusions can be found in ~\refse{sec:conclusions}.

\section{Model set-up}
\label{model_setup}

The majority of literature exploring the phenomenology of excited fermions typically operates under the assumption that these fermions possess $I_W=1/2$ weak isospin. However, one can introduce the higher isospin multiplets with $I_W=1,3/2$~\cite{Baur:1987ga, Baur:1989kv, Boudjema:1992em}. The  coupling of these  excited leptons to the gauge bosons  is given by the $SU(2) \times U(1)$ invariant (and CP conserving), effective Lagrangian \cite{Gonzalez-Garcia:1996oxz}:
\begin{eqnarray}
{\cal L}_{FF} &=& 
- \bar{\Psi}^\ast \Biggl[ \left(g \frac{\tau^i}{2} \gamma^\mu W^i_{\mu} + 
g' \frac{Y}{2}  \gamma^\mu B_\mu \right) \Biggr.\nonumber \\&\phantom{=}& + \Biggl.
\left(\frac{g k_2}{2 \Lambda} \frac{\tau^i}{2} \sigma^{\mu\nu}
\partial_\mu W^i_{\nu} + 
\frac{g' k_1}{2 \Lambda}  \frac{Y}{2} 
\sigma^{\mu\nu}\partial_\mu  B_\nu \right) 
\Biggr] \Psi^\ast\nonumber\\
\label{l:ee:0}
\end{eqnarray}
where, we represent the excited multiplet as $\Psi$, with its particle composition detailed in ~\refta{Lep-Multiplets}. The gauge coupling constants for $SU(2)$ and $U(1)$ are denoted as $g$ and $g'$, respectively, while $k_1$ and $k_2$ stand for dimensionless couplings. The constant $\Lambda$ denotes the compositeness scale.
In terms of the physical gauge fields, this can be written as: 
\begin{equation}
{\cal L}_{FF} = - \!\!\! \sum_{V=\gamma,Z,W} \bar{F} 
(A_{VFF} \gamma^\mu V_\mu + K_{VFF} \sigma^{\mu\nu} \partial_\mu V_\nu) F\; .
\label{l:ee}
\end{equation}
where $F$ denotes a generic excited fermion field appearing in the multiplet in ~\refta{Lep-Multiplets}.

\begin{table*}[t]
\begin{center}
\begin{tabular}{cccccl}
$I_{W}$ & Multiplet & $Q$ & $Y$ & Couple to & Couple through\\\hline
$0$ & $%
\begin{pmatrix}
E^{-}%
\end{pmatrix}
$ & $%
\begin{array}
[c]{c}%
-1
\end{array}
$ & $%
\begin{array}
[c]{c}%
-2
\end{array}
$ & $e_{R}$ & $B^{\mu}$\\\hline
$\frac{1}{2}$ & $%
\begin{pmatrix}
E^{0}\\
E^{-}%
\end{pmatrix}
$ & $%
\begin{array}
[c]{c}%
0\\
-1
\end{array}
$ & $%
\begin{array}
[c]{c}%
-1
\end{array}
$ & $%
\begin{pmatrix}
\nu_{e}\\
e
\end{pmatrix}
_{L}$ & $B^{\mu},W^{\mu}$\\\hline
$1$ & $%
\begin{pmatrix}
E^{0}\\
E^{-}\\
E^{--}%
\end{pmatrix}
$ & $%
\begin{array}
[c]{c}%
0\\
-1\\
-2
\end{array}
$ & $%
\begin{array}
[c]{c}%
-2
\end{array}
$ & $e_{R}$ & $W^{\mu}$\\\hline
$\frac{3}{2}$ & $%
\begin{pmatrix}
E^{+}\\
E^{0}\\
E^{-}\\
E^{--}%
\end{pmatrix}
$ & $%
\begin{array}
[c]{c}%
+1\\
0\\
-1\\
-2
\end{array}
$ & $%
\begin{array}
[c]{c}%
-1
\end{array}
$ & $%
\begin{pmatrix}
\nu_{e}\\
e
\end{pmatrix}
_{L}$ & $W^{\mu}$\\
\end{tabular}
\end{center}
\caption[Lepton multiplets]{Lepton multiplets for $I_W=0,1/2,1,3/2$
, their charge $Q$, hypercharge $Y$ and the fields through which they couple to ordinary leptons.}
\label{Lep-Multiplets}
\end{table*}
The higher multiplets include states with exotic charge like doubly charged leptons. The couplings involving $I_W=1$ excited fermions were calculated in ~\cite{Rehman:2020ana} which we reproduce here for completeness.
The couplings $A_{VFF}$ are given by:
\begin{equation}
\begin{array}{ll}
A_{\gamma E^{-}E^{-}} =  - e &
\; , \;\;\;\;
A_{\gamma E^{0}E^{0}} =   0 \\
A_{\gamma E^{--}E^{--}} = -2e  &
\; , \;\;\;\; 
A_{ZE^{0}E^{0}} =  \displaystyle\frac{e}{ s_W  c_W}
\\
A_{ZE^{-}E^{-}} = \displaystyle  \frac{ e s_W }{ c_W} &
\; , \;\;\;\; 
A_{ZE^{--}E^{--}} = \displaystyle\frac{ -e (1- 2 s^2_W) }{s_W  c_W} \\ 
A_{WE^{0}E^{-}} = \displaystyle\frac{ e}{ s_W}  &
\; , \;\;\;\; 
A_{WE^{-}E^{--}} =  \displaystyle\frac{e }{s_W}\\
A_{W E^{0}E^{--}} = 0  &
\end{array}
\label{AV}
\end{equation}
where $e$ represents the electric charge and $c_W$ ($s_W$) are the cosine (sin) of the weak mixing angle $\theta_W$. The couplings  $K_{VFF}$ are given by
\begin{equation}
\begin{array}{ll}
K_{\gamma E^{0}E^{0}} =
- \frac{\displaystyle e}{\displaystyle 2 \Lambda} (k_2 - k_1) &
,\;
K_{\gamma E^{-}E^{-}} =  
\frac{\displaystyle e}{\displaystyle 2 \Lambda} k_1 \\
K_{\gamma E^{--}E^{--}} = 
- \frac{\displaystyle e}{\displaystyle 2 \Lambda}
(k_2 + k_1)  &
,\;
K_{WE^{0}E^{-}} =  
\frac{\displaystyle e k_2}{\displaystyle 2 \Lambda s_W} \\
K_{WE^{-}E^{--}}= 
\frac{\displaystyle e}{\displaystyle 2 \Lambda s_W} &
,\;
K_{Z E^{-}E^{-}} = 
 \frac{\displaystyle e k_1 s_W}{\displaystyle 2 \Lambda c_W} \\ 
 K_{Z E^{0}E^{0}} =  
\frac{\displaystyle e ( k_1 s^2_W +  k_2 c^2_W)}{\displaystyle 2 \Lambda c_W s_W} & \;,\\
K_{Z E^{--}E^{--}} =  
\frac{\displaystyle e ( k_1 s^2_W -  k_2 c^2_W)}{\displaystyle 2 \Lambda c_W s_W} 
\end{array}
\label{KV}
\end{equation}
The $SU(2)\times U(1)$ invariant dimension-five effective Lagrangian that describe
the coupling of the excited fermions to the usual fermions can be written as \cite{Hagiwara:1985wt}
\begin{equation}
{\cal L}_{Ff} = - \frac{1}{2 \Lambda} {\bar \Psi^\ast} \sigma^{\mu\nu}
\left(g f \frac{\tau^i}{2} W^i_{\mu\nu} + 
      g' f^{\prime} \frac{Y}{2} B_{\mu\nu}\right) \psi_L 
+ \; \text{h. c.} ,
\label{l:eu:0}
\end{equation}
where $\sigma_{\mu\nu} = (i/2)[\gamma_\mu,
\gamma_\nu]$ and the dimensionless factors $f$ and $f^{\prime}$ are presumed to be of approximately the same magnitude, around unity, and are linked to the $SU(2)$ and $U(1)$ coupling constants respectively.  At tree-level, the couplings $g$ and $g'$ can  be
expressed in terms  of the electric charge, $e$, and the Weinberg
angle, $\theta_W$, as  $g=e/\sin\theta_W$ and
$g'=e/\cos\theta_W$. In terms of the physical fields, the Lagrangian (\ref{l:eu:0})
becomes
\begin{eqnarray}
{\cal L}_{Ff} &=& - \sum_{V=\gamma,Z,W} 
C_{VFf} \bar{F} \sigma^{\mu\nu} (1 - \gamma_5) f \partial_\mu V_\nu \nonumber \\
&\phantom{=}&- i \sum_{V=\gamma,Z} D_{VFf} \bar{F} \sigma^{\mu\nu} (1 - \gamma_5) f
W_\mu V_\nu +  \text{h.\! c.},
\label{l:eu}
\end{eqnarray}
where $F$ are the excited fermion states, $f$ the ordinary (SM) fermions  and $V=\gamma, Z,W$ are the physical vector boson fields. The non-abelian structure of (\ref{l:eu:0}) introduces a  quartic contact interaction,
such as the second term in the r.h.s. of Eq.\ (\ref{l:eu}). In this
equation, we have omitted terms containing two $W$ bosons, which
do not play any role in our calculations.

The interactions between excited fermions and SM fermions within the framework of higher weak isospin multiplets (triplet, \( I_W = 1 \), and quadruplet, \( I_W = 3/2 \)) were analyzed in~\citere{Pancheri:1984sm}. As demonstrated there, \refeq{CV} can be directly obtained from the effective Lagrangian by going from the gauge-interaction basis to the physical field basis.
For the case of an $I_W=1$ excited lepton triplet, the couplings $C_{VFf}$ and $D_{VFF}$ can be written as
\begin{equation}
\begin{array}{ll}
C_{\gamma E^{-} e} =  - \frac{\displaystyle e }{\displaystyle \Lambda} f_{1}\ 
&
\;\; , \;\;\;\;
C_{Z E^{-} e} =  
- \frac{\displaystyle e c_W}{\displaystyle \Lambda s_W} f_{1}\ \\
C_{W E^{0} e} =  
\frac{\displaystyle e }{\displaystyle \Lambda s_W} f_1\ &
\;\; , \;\;\;\;
C_{W E^{--} e} =    
\frac{\displaystyle e }{\displaystyle \Lambda s_W} f_1\
\\
& 
\end{array}
\label{CV}
\end{equation}
Here, $f_1$ represents the dimensionless parameter linked to the $SU(2)_L$ coupling constant for the triplet. The quartic interaction coupling constants, $D_{VFf}$, are given by
\begin{equation}
\begin{array}{l}
D_{\gamma E^{0} e} = - D_{\gamma E^{--} e}=   
 \frac{\displaystyle -e^2 }{\displaystyle 4 \Lambda s_W } f_1\ \\
D_{Z E^{0} e} = - D_{Z E^{--} e} = 
 \frac{\displaystyle e^2 c_W}
{\displaystyle 4 \Lambda s^2_W } f_1\ \\
D_{W E^{-} e} =  
 \frac{\displaystyle - e^2 }
{\displaystyle 4 \Lambda s^2_W } f_1\
\end{array}
\label{DV}
\end{equation}


\section{Excited fermion contributions to $(g-2)_{\mu}$}
\label{Aresults}

\begin{figure}[htb!]
\centering
\includegraphics[width=14cm,keepaspectratio]{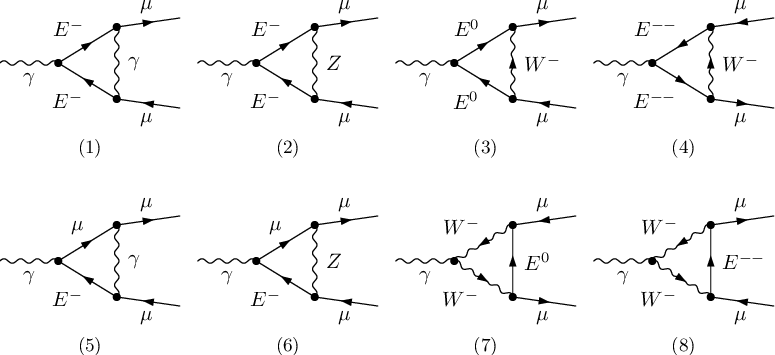}
\caption{Vertex diagrams for the excited lepton contributions to $\Delta a_{\mu}^{\rm Exc}$.}   
\label{fig:vertex}
\end{figure}

\begin{figure}[htb!]
\centering
\includegraphics[width=14cm,keepaspectratio]{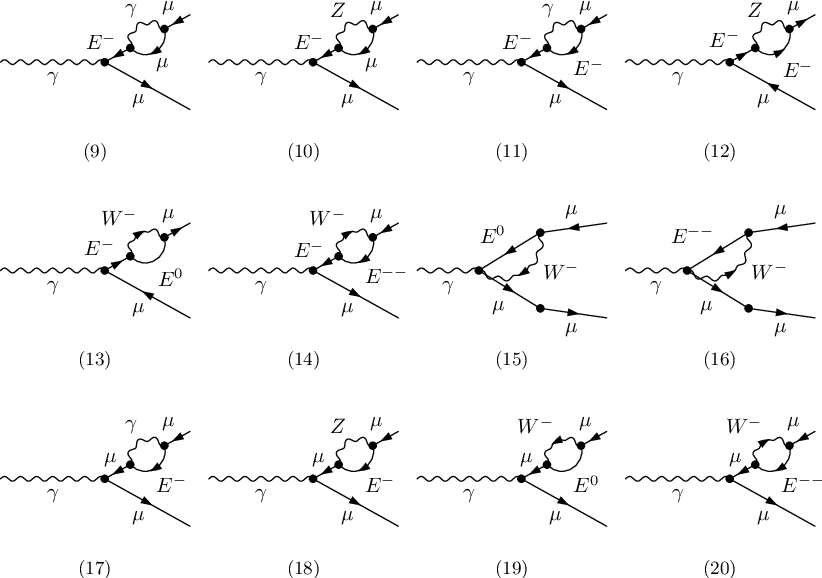}
\caption{Self-energy diagrams for the excited lepton contributions to $\Delta a_{\mu}^{\rm Exc}$.}   
\label{fig:self}
\end{figure}

The contributions of excited leptons to $(g-2)_{\mu}$ in the case of isospin doublets have already been studied in \citere{Gonzalez-Garcia:1996wkh}, where the evaluation of one-loop contributions to the magnetic form factors was performed using dimensional regularization technique. Utilizing the generic results presented in that study, we have derived the contributions of excited lepton triplets for the couplings specified in Eqs.~(\ref{AV}), (\ref{KV}), (\ref{CV}), and (\ref{DV}). The vertex diagrams, encompassing contributions from excited lepton triplets in loops, are shown in \reffi{fig:vertex}, while the self-energy diagrams are shown in \reffi{fig:self}. Each Feynman diagram is associated with an excited fermion contribution denoted as $\Delta a_{\mu}^{i}$, where $i=1,2,3......20$ corresponds to the respective diagram number. Employing this definition, we have

\begin{equation}
\Delta a_{\mu}^{\rm Exc}=\sum\limits_{i=1}^{20} \Delta a_{\mu}^{i}
\end{equation}

In the limit where $q^2 \ll M^2$ and considering the first order in $R_Q$ with $R_Q=q^2/M^2$ and $R_V=M_{V}^{2}/M^2$, the corrections originating from the vertex diagrams shown in \reffi{fig:vertex} can be expressed as

\begin{align}
\Delta a_{\mu}^{1}  & =\frac{1}{36\pi}\frac{m_{\mu}^{2}\alpha}{\Lambda^{2}}%
f^{2}_{1}\left[  120-108(\frac{M}{2\Lambda})k_{1}+72(-1+(\frac{M}{2\Lambda}%
)k_{1})\log\frac{\Lambda^{2}}{M^{2}}\right]  \\ \label{AnaResStart}
\Delta a_{\mu}^{2}  & =\frac{1}{36\pi}\frac{m_{\mu}^{2}\alpha}{\Lambda^{2}}f^{2}_{1}%
\frac{C_{W}^{2}}{S_{W}^{2}}\left[
\begin{array}
[c]{c}%
6(20+9R_{Z})-36(\frac{M}{2\Lambda}k_{1})(3+R_{Z})\\
+72(-1+(\frac{M}{2\Lambda}k_{1}))\log\frac{\Lambda^{2}}{M^{2}}%
\end{array}
\right]  \\
\Delta a_{\mu}^{3}  & =\frac{1}{36\pi}\frac{m_{\mu}^{2}\alpha}{\Lambda^{2}S_{W}^{2}%
}f^{2}_{1}\left[  36(\frac{M}{2\Lambda})(k_{2}-k_{1})(3+R_{W})-72(\frac
{M}{2\Lambda})(k_{2}-k_{1})\log\frac{\Lambda^{2}}{M^{2}}\right]  \\
\Delta a_{\mu}^{4}  & =\frac{1}{36\pi}\frac{m_{\mu}^{2}\alpha}{\Lambda^{2}S_{W}^{2}%
}f^{2}_{1}\left[
\begin{array}
[c]{c}%
12(20+9R_{W})+36(\frac{M}{2\Lambda})(k_{2}+k_{1})(3+R_{W})\\
-72(2+(\frac{M}{2\Lambda})(k_{2}+k_{1}))\log\frac{\Lambda^{2}}{M^{2}}%
\end{array}
\right]  \\
\Delta a_{\mu}^{5}  & =\frac{1}{2\pi}\frac{m_{\mu}^{2}\alpha}{\Lambda
^{2}}f^{2}_{1}\left[  1+2\log\frac{\Lambda^{2}}{M^{2}}\right]  \\
\Delta a_{\mu}^{6}  & =\frac{-1}{24\pi}(\frac{m_{\mu}^{2}\alpha}%
{\Lambda^{2}S_{W}})f^{2}_{1}\left[
\begin{array}
[c]{c}%
(4S_{W}^{2}-1)(3+6R_{Z}+12R_{Z}\log R_{Z}+6\log\frac{\Lambda^{2}}{M^{2}})\\
-(39+6R_{Z}+12R_{Z}\log R_{Z}+6\log\frac{\Lambda^{2}}{M^{2}})
\end{array}
\right]  \\
\Delta a_{\mu}^{7}  & =\Delta a_{\mu}^{8}=\frac{1}{36\pi}\frac{m_{\mu}^{2}\alpha}%
{\Lambda^{2}S_{W}}f^{2}_{1}\left[  -79-120R_{W}+42\log\frac{\Lambda^{2}}{M^{2}%
}\right]
\end{align}
where $m_\mu$ represents the mass of the muon and $\alpha$ denotes the fine-structure constant.

Under the previously specified limit, the approximate corrections to $\Delta a_{\mu}^{\rm Exc}$, considering the first order in $R_Q$, arising from the self-energy diagrams presented in \reffi{fig:self}, can be written as

\begin{align}
\Delta a_{\mu}^{9}  & =\frac{2}{\pi}\frac{m_{\mu}^{2}\alpha}%
{\Lambda^{2}}f^{2}_{1}\left(  \frac{\Lambda^{2}}{M^{2}}\right)  \\ 
\Delta a_{\mu}^{10}  & =\frac{-1}{\pi}\frac{m_{\mu}^{2}\alpha}%
{\Lambda^{2}S_{W}}f^{2}_{1}(2S_{W}^{2}-1)\left(  2R_{Z}+3R_{Z}\log R_{Z}%
-3R_{Z}\log\frac{\Lambda^{2}}{M^{2}}+\frac{\Lambda^{2}}{M^{2}}\right)  \\
\Delta a_{\mu}^{11}  & =\frac{1}{2\pi}\frac{m_{\mu}^{2}\alpha}%
{\Lambda^{2}}f^{2}_{1}\left[
\begin{array}
[c]{c}%
15-22(\frac{M}{2\Lambda}k_{1})+6\left(  -3+4(\frac{M}{2\Lambda}k_{1})\right)
\log\frac{\Lambda^{2}}{M^{2}}\\
-4\frac{\Lambda^{2}}{M^{2}}\left(  -1+(\frac{M}{2\Lambda}k_{1})\right)
\end{array}
\right]  \\
\Delta a_{\mu}^{12}  & =\frac{-1}{2\pi}\frac{m_{\mu}^{2}\alpha}%
{\Lambda^{2}}f^{2}_{1}\left[
\begin{array}
[c]{c}%
15+14R_{Z}+(\frac{M}{2\Lambda}k_{1})(22+21R_{Z})\\
-6\left(  3+2R_{Z}+(\frac{M}{2\Lambda}k_{1})(4+3R_{Z})\right)  \log
\frac{\Lambda^{2}}{M^{2}}\\
+4\frac{\Lambda^{2}}{M^{2}}\left(  1+(\frac{M}{2\Lambda}k_{1})\right)
\end{array}
\right]  \\
\Delta a_{\mu}^{13}  & =\frac{1}{2\pi}\frac{m_{\mu}^{2}\alpha}{S_{W}%
^{2}\Lambda^{2}}f^{2}_{1}\left[
\begin{array}
[c]{c}%
15+14R_{W}+(\frac{M}{2\Lambda}k_{2})(22+21R_{W})\\
-6\left(  3+2R_{W}+(\frac{M}{2\Lambda}k_{2})(4+3R_{W})\right)  \log
\frac{\Lambda^{2}}{M^{2}}\\
+4\frac{\Lambda^{2}}{M^{2}}\left(  1+(\frac{M}{2\Lambda}k_{2})\right)
\end{array}
\right]  \\
\Delta a_{\mu}^{14}  & =\frac{1}{2\pi}\frac{m_{\mu}^{2}\alpha}{S_{W}%
^{2}\Lambda^{2}}f^{2}_{1}\left[
\begin{array}
[c]{c}%
15+14R_{W}+(\frac{M}{2\Lambda})(22+21R_{W})\\
-6\left(  3+2R_{W}+(\frac{M}{2\Lambda})(4+3R_{W})\right)  \log\frac
{\Lambda^{2}}{M^{2}}\\
+4\frac{\Lambda^{2}}{M^{2}}\left(  1+(\frac{M}{2\Lambda})\right)
\end{array}
\right]  \\
\Delta a_{\mu}^{15}  & =-\Delta a_{\mu}^{16}=\frac{1}{36\pi}%
\frac{m_{\mu}^{2}\alpha}{S_{W}^{2}\Lambda^{2}}f^{2}_{1}\left(  1+12R_{W}%
-6\log\frac{\Lambda^{2}}{M^{2}}\right)  \\
\Delta a_{\mu}^{17}  & =\Delta a_{\mu}^{18}=\Delta a_{\mu}^{19}=\Delta a_{\mu}^{20}=0 \label{AnaResEnd}
\end{align}

The results presented above are obtained with the assumption that the masses of the excited leptons are degenerate. 

\section{Numerical Results}
\label{Nresults}

\subsection{Experimental Status of $(g-2)_{\mu}$}
Considerable attention has been directed toward the longstanding anomaly observed in the value of $(g-2)_{\mu}$. The most recent assessment, undertaken by the Run II and Run III of the Muon $g-2$ collaboration at Fermilab \cite{Muong-2:2023cdq}, when combined with earlier findings from the same experiment \cite{Muong-2:2021ojo} and the Brookhaven E821 experiment \cite{Muong-2:2006rrc}, reveals a deviation of $5.1\sigma$ from the SM prediction\cite{Aoyama:2020ynm, Martin:2001st}.

\begin{align}
\Delta \alpha_{\mu}^{\text{WA}}=\alpha_{\mu}^{\rm exp}-\alpha_{\mu}^{\rm SM,~WA}= (24.9\pm 4.8) \times 10^{-10}.
\label{Eq:delAmu}
\end{align}
However, recent re-analysis of isospin-breaking (IB) corrections to
$e^{+}e^{-}$ and $\tau$-decay di-pion observables \cite{Miranda:2024ojv} has reduced the
discrepancy to the $2.7\sigma$ level:%
\begin{equation}
\Delta a_{\mu}^{\text{IB}} = a_{\mu}^{\text{exp}} - a_{\mu}^{\text{SM, IB}} =
(14.8_{-5.4}^{+5.1}) \times10^{-10}.
\end{equation}
Using the latest Budapest-Marseille-Wuppertal (BMW) calculations for the hadronic vacuum polarization\cite{Boccaletti:2024guq} and
light-by-light scattering contributions\cite{Fodor:2024jyn}, the discrepancy is further reduced to
less than $1\sigma$~\cite{Coutinho:2024zyp}:%

\begin{equation}
\Delta a_{\mu}^{\text{BMW}} = a_{\mu}^{\text{exp}} - a_{\mu}^{\text{SM, BMW}} =
(0.4 \pm4.2) \times10^{-10}.
\end{equation}

The value of $\Delta a_{\mu}^{\text{BMW}}$ is compatible with zero, indicating no discrepancy between the SM prediction and experimental results, leaving little to no room for new physics effects. In contrast, $\Delta \alpha_{\mu}^{\text{BMW}}$ requires moderate contributions from new physics, while $\Delta \alpha_{\mu}^{\text{WA}}$ requires even greater contributions. This highlights that the status of the $(g-2)_{\mu}$ discrepancy remains uncertain. However, a detailed analysis of new physics contributions to $(g-2)_{\mu}$ could provide valuable insights into this discrepancy and the associated parameter space of new physics.

\subsection{Input Parameters}

The essential parameters for defining the effective theory include the energy scale $\Lambda$, the mass range of exotic particles denoted by $M$, and the dimensionless factors $k_1$, $k_2$, $f^{\prime}$, $f$ and $f_1$. The factors $f^{\prime}$ and $f$ are involved in the doublet calculations, whereas the parameter $f_1$ pertains to triplets. For our numerical evaluation, we examined the value of $\Lambda$ within the interval $0 < \Lambda < 25 \tev$ and $M$ within the range $0 < M < 12 \tev$. However, $k_1$ and $k_2$ are set to unity since their influence on our calculations is negligible. Yet, we've observed that while $f^{\prime}$, $f$ and $f_1$ are inconsequential for $\Delta \rho$, they can significantly affect the computation of $\Delta \alpha_{\mu}$. We investigate two cases for $f^{\prime}$ and $f$: one assumes equal values, denoted as $f^{\prime}=f$, while the other examines various combinations for $f^{\prime}$ and $f$.

Considering that excited fermion masses typically emerge above the Electroweak Symmetry Breaking (EWSB) scale, we can presume that they have similar magnitudes. However, our previous work \cite{Rehman:2020ana} highlighted $\Delta \rho$'s sensitivity to doublet and triplet mass non-degeneracy. Although $\Delta \alpha_{\mu}$ is not affected by this, it's advantageous to allow for such possibilities to integrate constraints from both $\Delta \alpha_{\mu}$ and $\Delta \rho$. Hence, we incorporate the potential for slight mass differences among exotic fermions within the same isospin multiplet, which may arise from SU(2) breaking contributions such as interactions with exotic Higgs bosons.

\subsection{Contributions to $(g-2)_{\mu}$ from degenerate Fermion Masses with $f^{\prime}=f$}
\label{degen-mass}

\subsubsection{Doublet Contributions}

\begin{figure}[htb!]
\begin{center}
\psfig{file=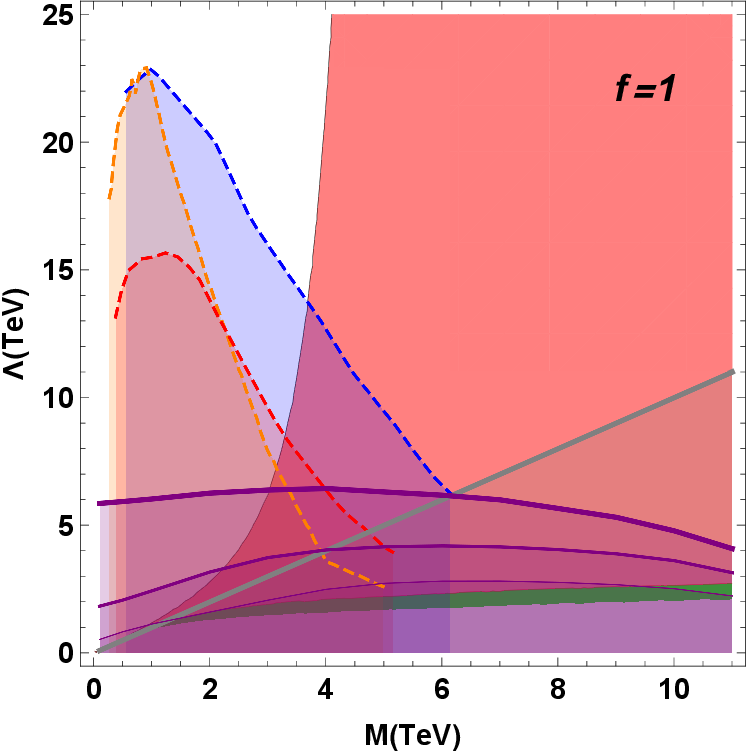 ,scale=0.50,angle=0,clip=}
\psfig{file=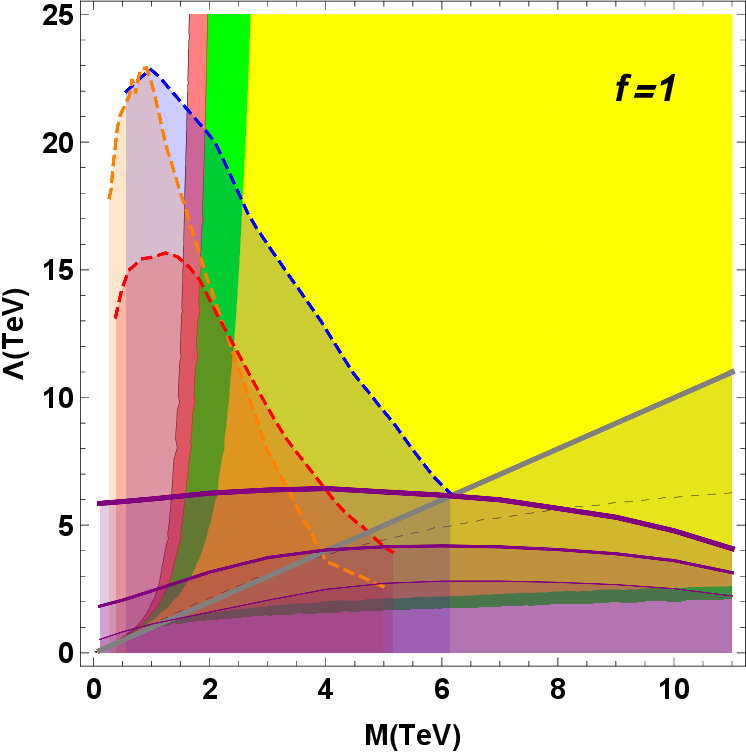  ,scale=0.50,angle=0,clip=}
\end{center}
\caption{$\Delta a_{\mu}^{\rm IB}$ (left plot ) and $\Delta a_{\mu}^{\rm BMW}$ (right plot) in $(M, \Lambda)$ plane for the case of excited lepton doublet. The dashed black line represents the central value of $\Delta a_{\mu}^{\rm IB,~BMW}$, while the yellow, green, and pink regions denote the $1\sigma$, $2\sigma$, and $3\sigma$ regions, respectively. The shaded area under the gray line is excluded by the $M > \Lambda$ constraint. Unitarity bound (purple lines)\cite{Biondini:2019tcc} and exclusion limits (blue, orange and red dashed lines) \cite{CMS:2018sfq, CMS:2022chw, ATLAS:2023kek} from CMS and ATLAS experiments for charged leptons searches with two different final states are also shown for comparison.} 
\label{fig:degen-Doub-F1}
\end{figure}

Utilizing the formulas provided in Eqs.~(\ref{AnaResStart}-\ref{AnaResEnd}), we can estimate $\Delta a_{\mu}^{\rm Exc}$. Our results for doublet contributions with $f^{\prime}=f=1$ are presented in \reffi{fig:degen-Doub-F1}, showcasing contours of $\Delta a_{\mu}^{\rm IB}$ (left plot ) and $\Delta a_{\mu}^{\rm BMW}$ (right plot) in the $(M,\Lambda)$ plane. The dashed black line represents the central value of $\Delta a_{\mu}^{\rm IB, ~BMW}$, while the yellow, green, and pink regions denote the $1\sigma$, $2\sigma$, and $3\sigma$ regions, respectively. Additionally, these plots incorporate experimental findings, indicating exclusion regions in the $(M,\Lambda)$ plane with a 95$\%$ confidence level \cite{CMS:2018sfq, CMS:2022chw, ATLAS:2023kek}. The dashed blue line corresponds to the recent CMS search for an excited lepton decaying to two muons and two jets via contact interaction, with a total integrated luminosity of $139$ $\rm{fb^{-1}}$, the orange line corresponds to the search in the $\ell \ell \gamma$ channel with an integrated luminosity of $35.9$ $\rm{fb^{-1}}$ and the red line corresponds to recent ATLAS search in the $\tau \tau$jj channel with an integrated luminosity of $139$ $\rm{fb^{-1}}$. Perturbative unitarity bounds for a composite fermion model were explored in \cite{Biondini:2019tcc}, with the purple line representing these bounds, and the decreasing thickness of the line corresponding to 100$\%$, 95$\%$, and 50$\%$ event fractions, respectively, that satisfy unitarity bounds. The shaded area is excluded due to unitarity bounds (purple lines) and direct searches at the LHC (blue, orange and red lines).

The $\Delta a_{\mu}^{\rm WA}$ within $3 \sigma$ range cannot be reached for $f = 1$,  and is therefore not displayed. Similarly, the central values of $\Delta a_{\mu}^{\rm IB}$ and $\Delta a_{\mu}^{\rm BMW}$ lie within regions already excluded by experimental constraints as can be seen in \reffi{fig:degen-Doub-F1}. However, the $3\sigma$ region for $\Delta a_{\mu}^{\rm IB}$ and the $1\sigma$ region for $\Delta a_{\mu}^{\rm BMW}$ remain viable.

\begin{figure}[htb!]
\begin{center}
\psfig{file=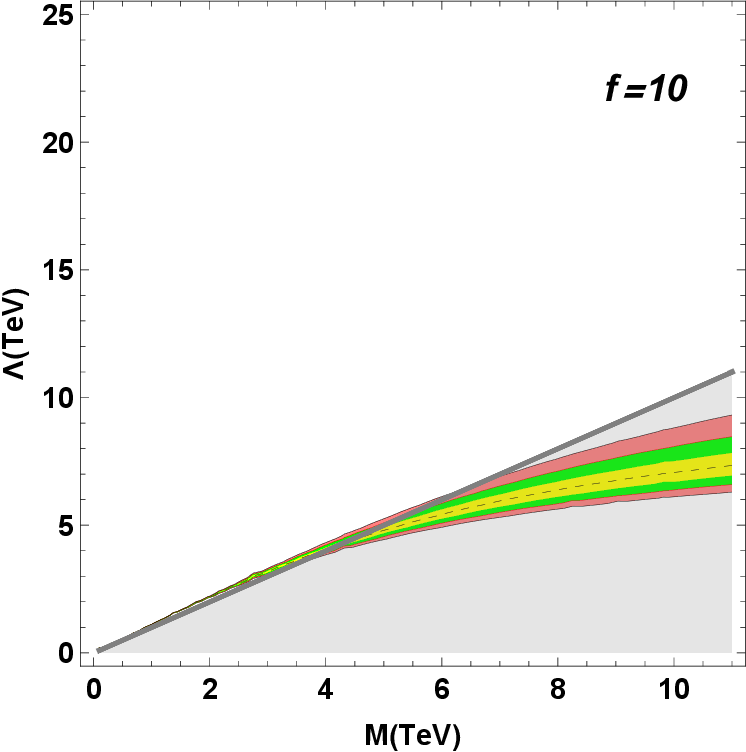  ,scale=0.40,angle=0,clip=}
\psfig{file=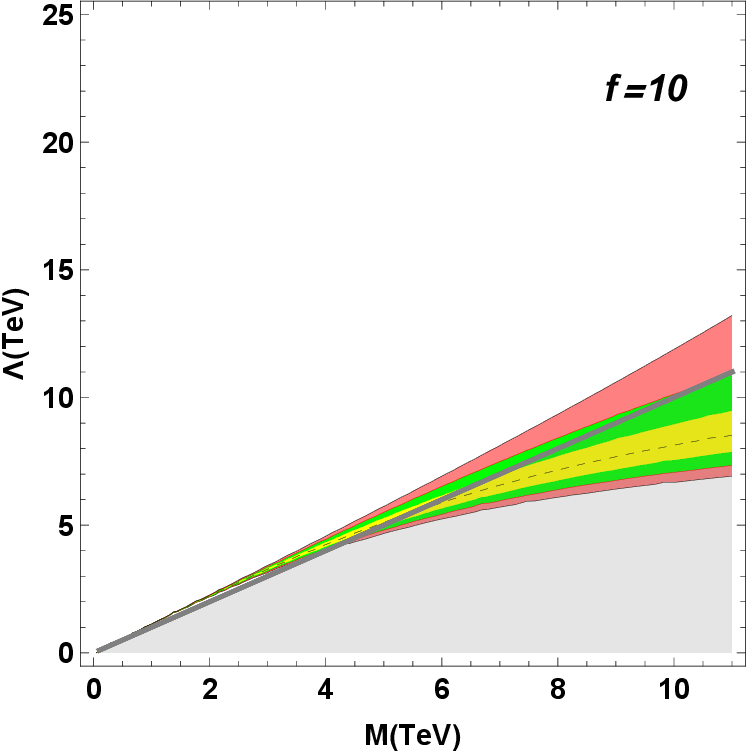 ,scale=0.40,angle=0,clip=}
\psfig{file=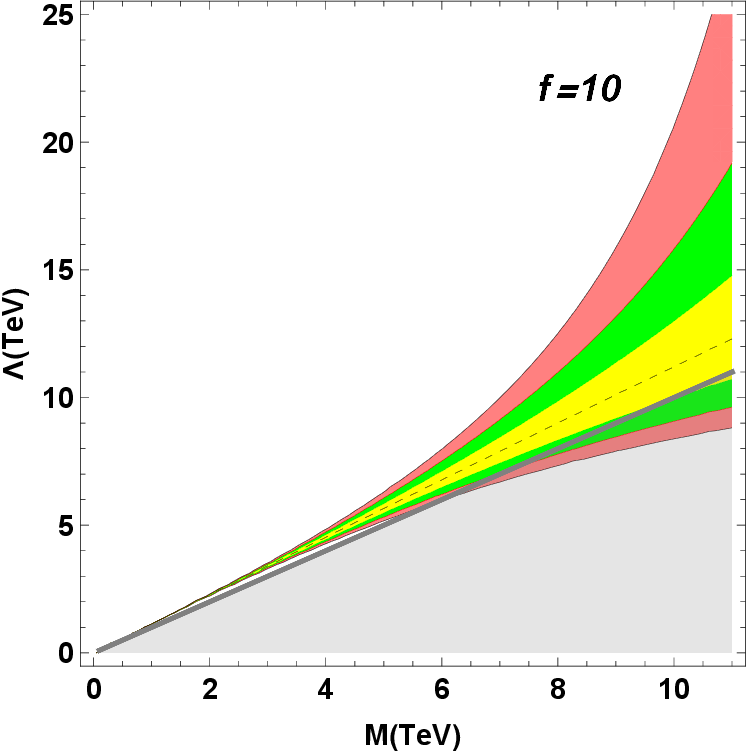  ,scale=0.40,angle=0,clip=}\\
\psfig{file=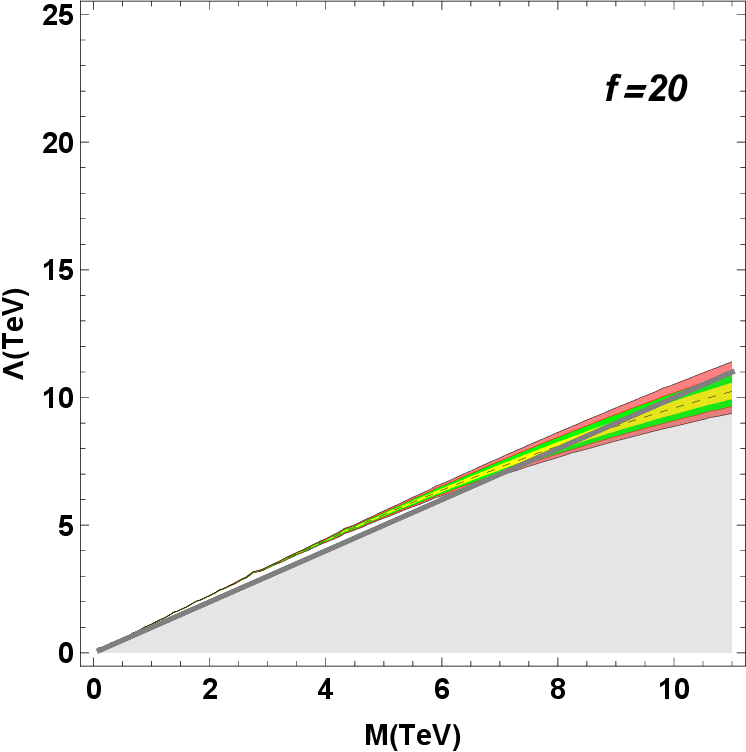  ,scale=0.40,angle=0,clip=}
\psfig{file=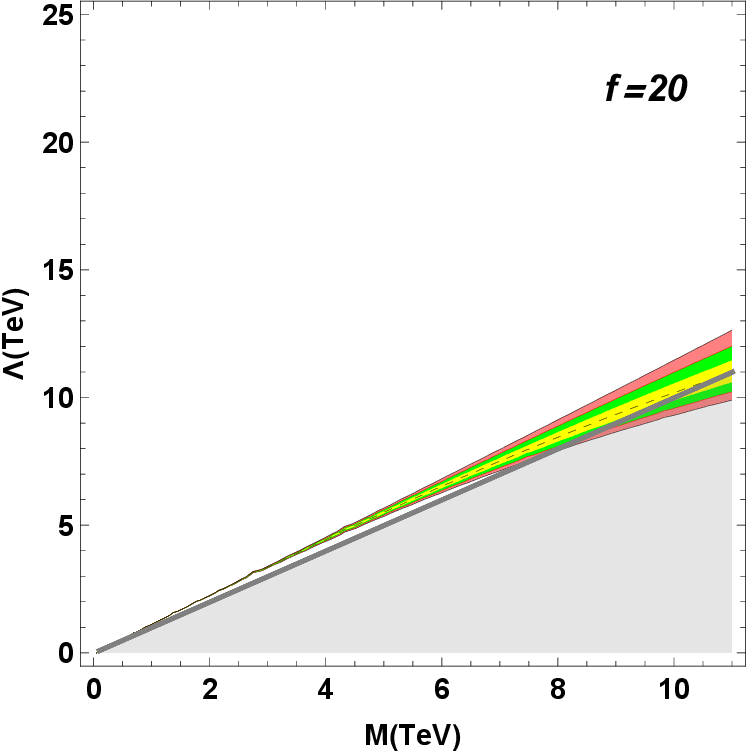 ,scale=0.40,angle=0,clip=}
\psfig{file=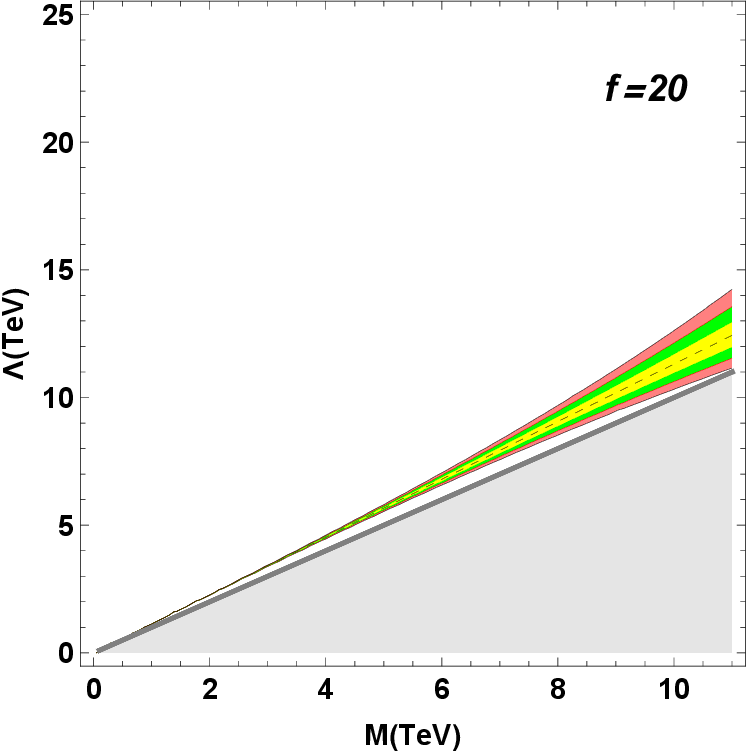  ,scale=0.40,angle=0,clip=}
\end{center}
\caption{$\Delta a_{\mu}^{\rm WA}$ (left plot), $\Delta a_{\mu}^{\rm IB}$ (middle plot ) and $\Delta a_{\mu}^{\rm BMW}$ (right plot) in $(M, \Lambda)$ plane for the case of excited lepton doublet. The dashed black line represents the central value of $\Delta a_{\mu}^{\rm WA,~IB,~BMW}$, while the yellow, green, and pink regions denote the $1\sigma$, $2\sigma$, and $3\sigma$ regions, respectively. The shaded area under the gray line is excluded by the $M > \Lambda$ constraint.} 
\label{fig:degen-Doub-F10-F20}
\end{figure}

The contours of $\Delta a_{\mu}^{\rm WA,~IB,~BMW}$ for the weight factors $f = 10$ (upper row) and $f = 20$ (lower row) are shown in \reffi{fig:degen-Doub-F10-F20}. The lines and color coding are consistent with those in the previous figure. However, unitarity and experimental constraints are not displayed here, as they are unavailable for higher values of $f$.

These plots highlight the significant sensitivity of $\Delta a_{\mu}^{\rm WA,~IB,~BMW}$ to the chosen value of the weight factor $f$. For smaller values, such as $f = 10$, the doublet contributions are minimal and primarily fall within regions already excluded by the $M > \Lambda$ constraint, except for $\Delta a_{\mu}^{\rm BMW}$, which remains viable. In contrast, for larger values of $f$, such as $f = 20$, the required value of $\Delta a_{\mu}^{\rm WA,~IB,~BMW}$ can be achieved in regions still allowed by the $M > \Lambda$ constraint. However, the bands corresponding to the $1\sigma$, $2\sigma$, and $3\sigma$ regions become significantly narrower.

\subsubsection{Triplet Contributions}

\begin{figure}[htb!]
\begin{center}
\psfig{file=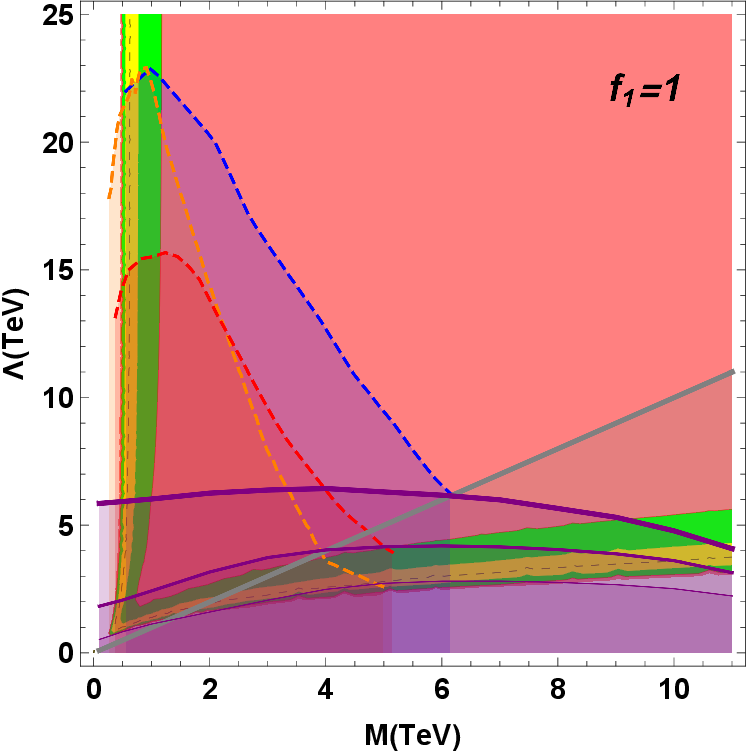 ,scale=0.50,angle=0,clip=}
\psfig{file=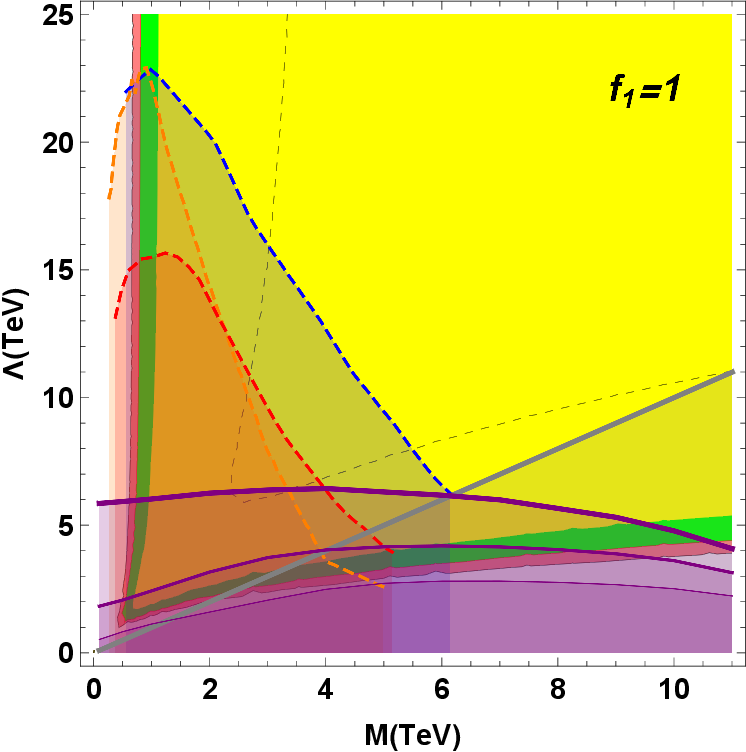  ,scale=0.50,angle=0,clip=}
\end{center}
\caption{$\Delta a_{\mu}^{\rm IB}$ (left plot ) and $\Delta a_{\mu}^{\rm BMW}$ (right plot) in $(M, \Lambda)$ plane for the case of excited lepton triplet. The lines and color coding is the same as in  \reffi{fig:degen-Doub-F1}.} 
\label{fig:degen-Trip-F1}
\end{figure}

Our results for the triplet contribution with the weight factor $f_1 = 1$ are presented in \reffi{fig:degen-Trip-F1}. The shading of the contours follows the same convention as in \reffi{fig:degen-Doub-F1}. Once again, the $3\sigma$ range of $\Delta a_{\mu}^{\rm WA}$ cannot be achieved, and therefore only the plots for $\Delta a_{\mu}^{\rm IB}$ and $\Delta a_{\mu}^{\rm BMW}$ are shown. The central value of $\Delta a_{\mu}^{\rm IB}$ lies in the already excluded region; however, the central value of $\Delta a_{\mu}^{\rm BMW}$ falls within the allowed region, as seen in the right plot of \reffi{fig:degen-Trip-F1}.

\begin{figure}[htb!]
\begin{center}
\psfig{file=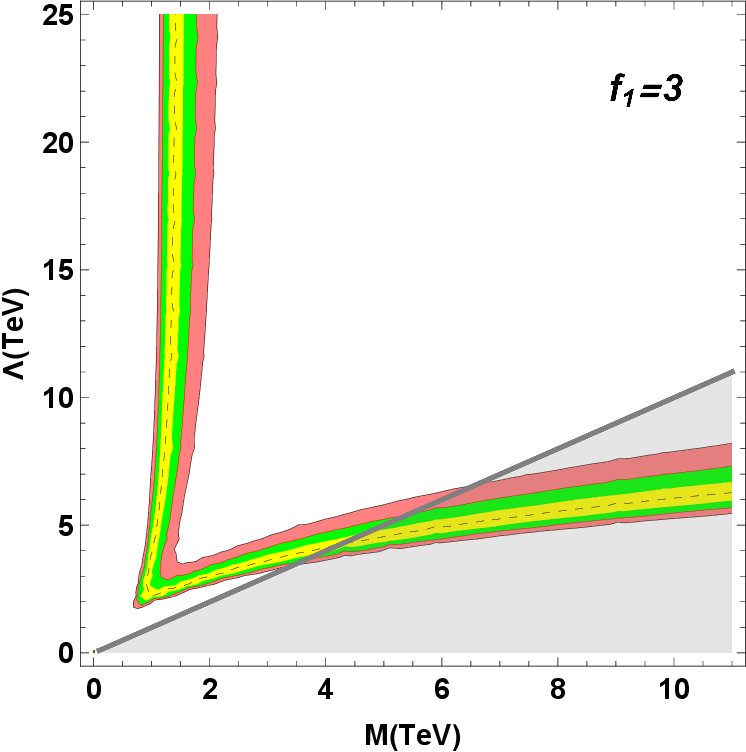  ,scale=0.40,angle=0,clip=}
\psfig{file=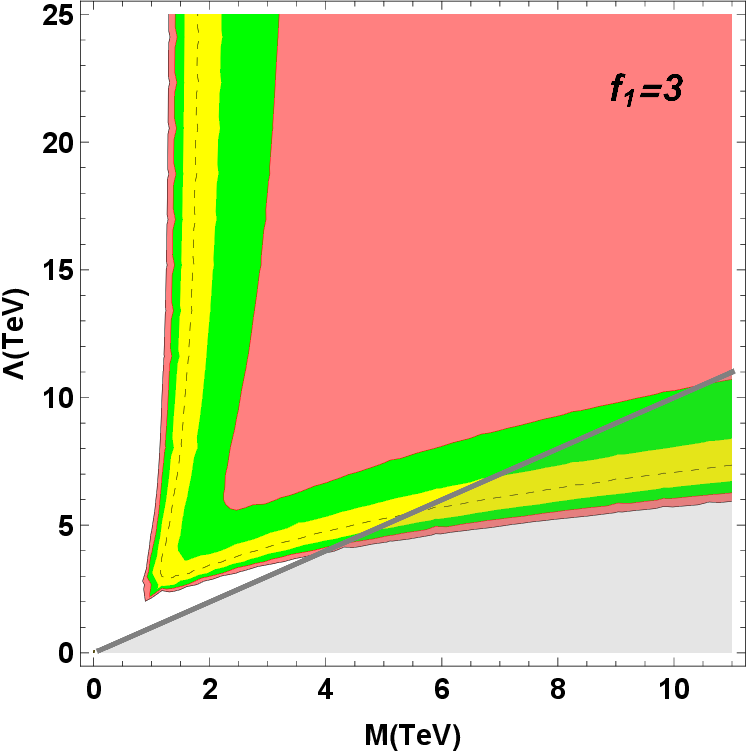 ,scale=0.40,angle=0,clip=}
\psfig{file=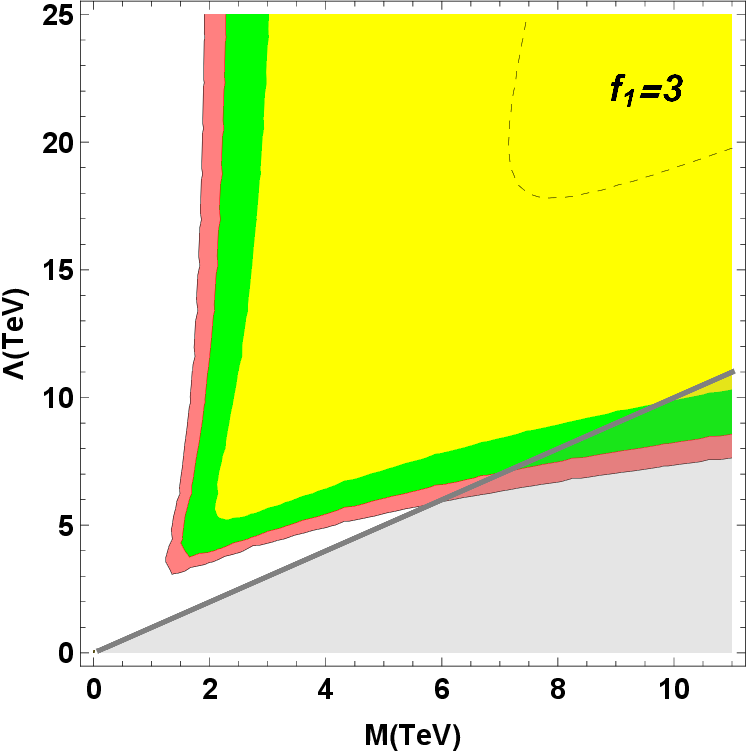  ,scale=0.40,angle=0,clip=}\\
\psfig{file=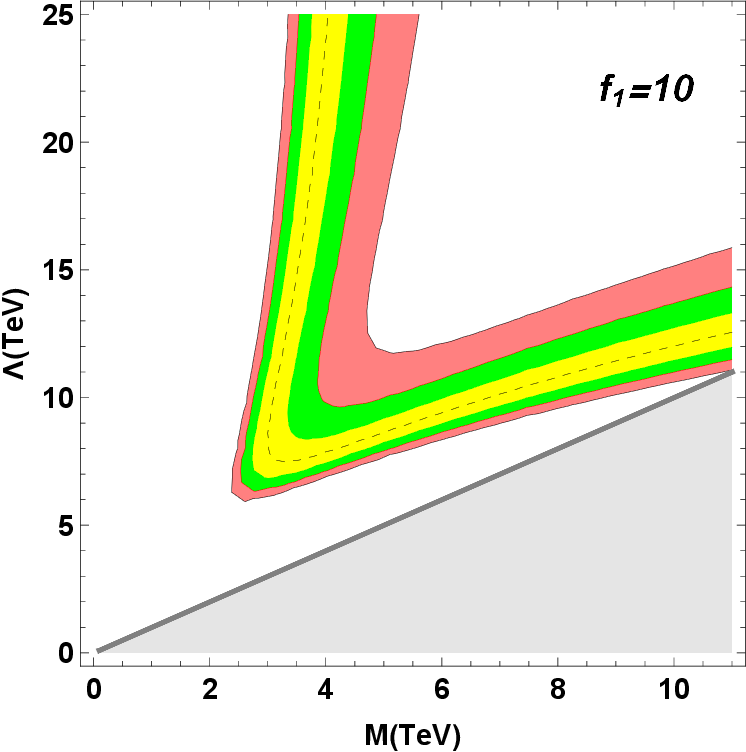  ,scale=0.40,angle=0,clip=}
\psfig{file=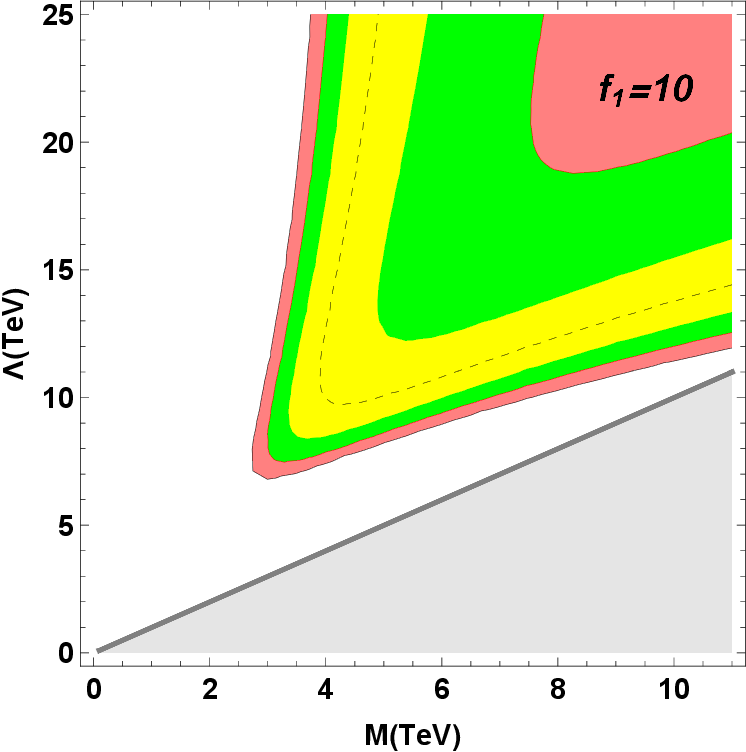 ,scale=0.40,angle=0,clip=}
\psfig{file=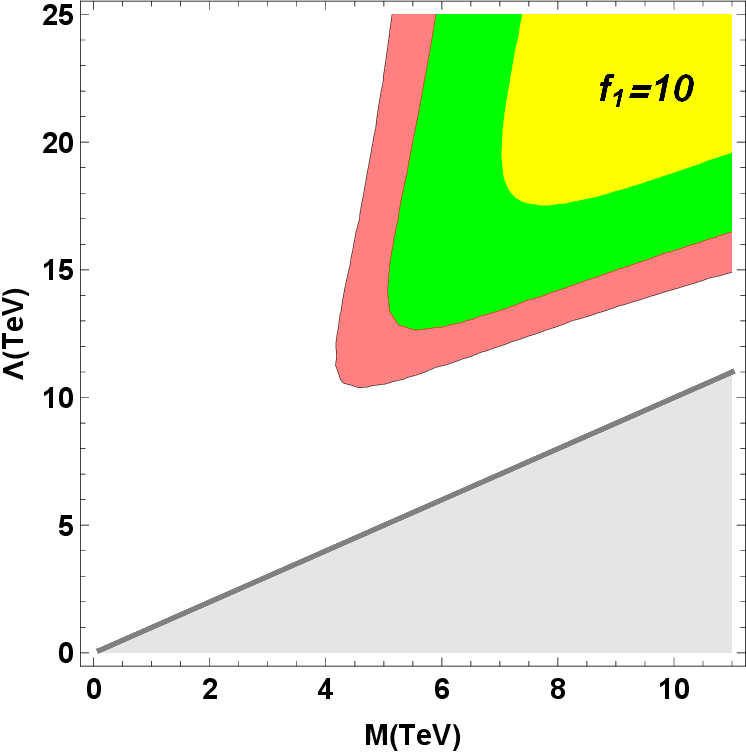  ,scale=0.40,angle=0,clip=}
\end{center}
\caption{$\Delta a_{\mu}^{\rm WA}$ (left plot), $\Delta a_{\mu}^{\rm IB}$ (middle plot ) and $\Delta a_{\mu}^{\rm BMW}$ (right plot) in $(M, \Lambda)$ plane for the case of excited lepton doublet. The dashed black line represents the central value of $\Delta a_{\mu}^{\rm WA,~IB,~BMW}$, while the yellow, green, and pink regions denote the $1\sigma$, $2\sigma$, and $3\sigma$ regions, respectively. The shaded area under the gray line is excluded by the $M > \Lambda$ constraint.} 
\label{fig:degen-Trip-F3-F10}
\end{figure}

In \reffi{fig:degen-Trip-F3-F10}, we present our results for $f_1 = 3$ and $f_1 = 10$. As shown in the upper row, for $f_1 = 3$, the $3\sigma$ region for $\Delta a_{\mu}^{\rm WA}$ is very narrow, whereas the $3\sigma$ regions for $\Delta a_{\mu}^{\rm IB}$ and $\Delta a_{\mu}^{\rm BMW}$ remain wide open. For larger values of $f_1$, specifically $f_1 = 10$, the required values of $\Delta a_{\mu}^{\rm WA,~IB,~BMW}$ can be achieved within regions that are still permitted. Notably, the triplet contributions are relatively more significant than the doublet contributions for the same value of $f_1$.

In summary, for both doublet and triplet contributions, $\Delta a_{\mu}^{\rm WA}$ is achievable only within a highly restricted region of the $(M, \Lambda)$ plane, particularly for smaller values of the weight factors $f$ and $f_1$. In contrast, $\Delta a_{\mu}^{\rm IB}$ and $\Delta a_{\mu}^{\rm BMW}$ can be explained relatively easily across a broader parameter space. This disparity highlights the inherent sensitivity of $\Delta a_{\mu}^{\rm WA}$ to the chosen weight factors and parameter values. Consequently, the stringent requirements for $\Delta a_{\mu}^{\rm WA}$ serve as an indirect but significant constraint on the viable parameter space, further refining the regions that can accommodate the experimental observations.

\subsection{Contributions to $(g-2)_{\mu}$ from degenerate Fermion Masses with $f^{\prime} \neq f$}

In the literature, it is usually assumed that the weight factors associated with $SU(2)$ and $U(1)$ couplings are equal i.e. $f^{\prime}=f$. In theory, they can indeed differ. In triplet calculations, only $f_1$ is involved, thus yielding unchanged results. Conversely, for doublets, both factors $f^{\prime}$ and $f$ hold significance. In this section, we explore the cases where $f^{\prime} \neq f$. 

\begin{figure}[htb!]
\begin{center}
\psfig{file=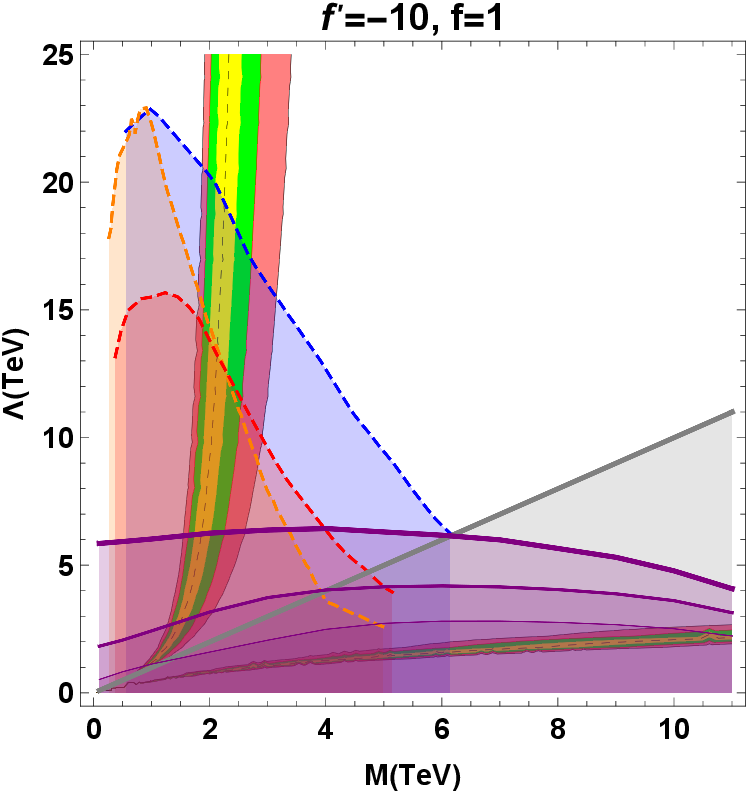  ,scale=0.40,angle=0,clip=}
\psfig{file=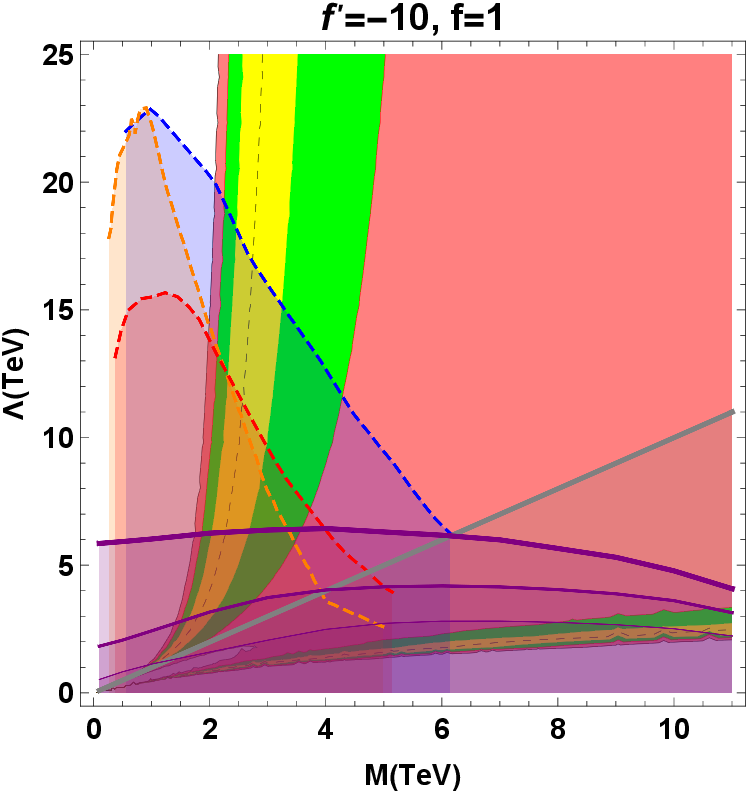  ,scale=0.40,angle=0,clip=}
\psfig{file=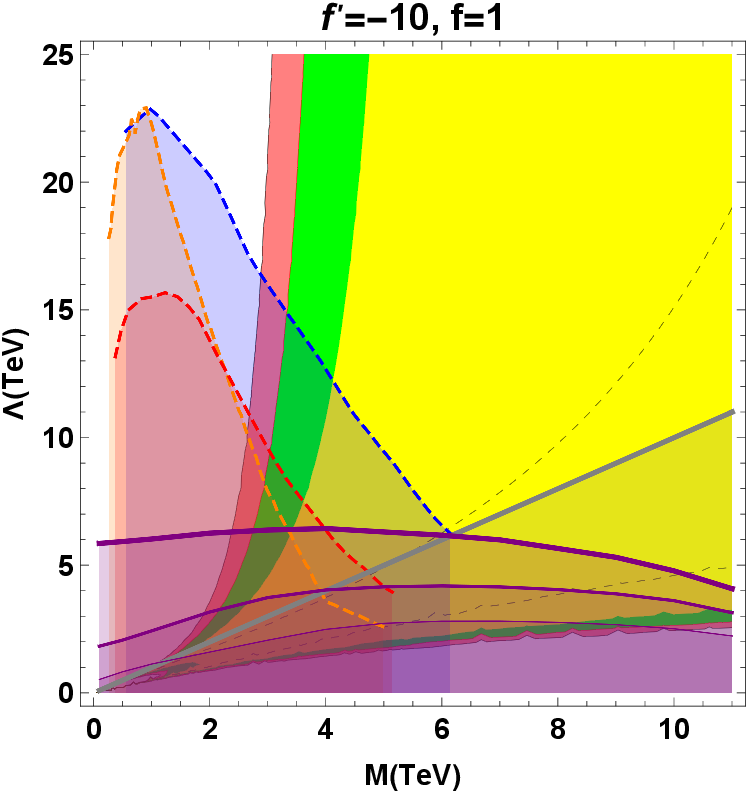  ,scale=0.40,angle=0,clip=} \\
\psfig{file=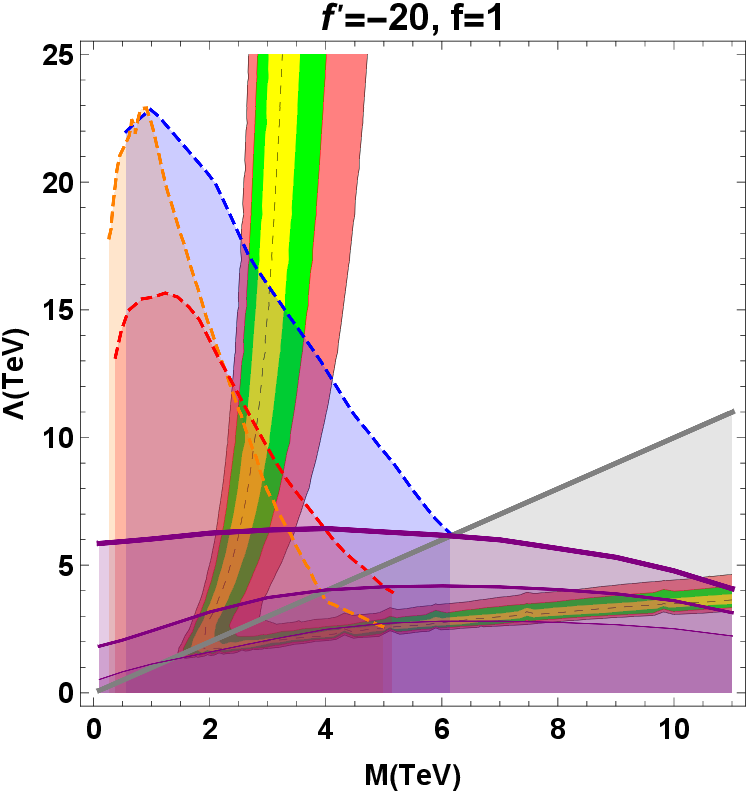 ,scale=0.40,angle=0,clip=}
\psfig{file=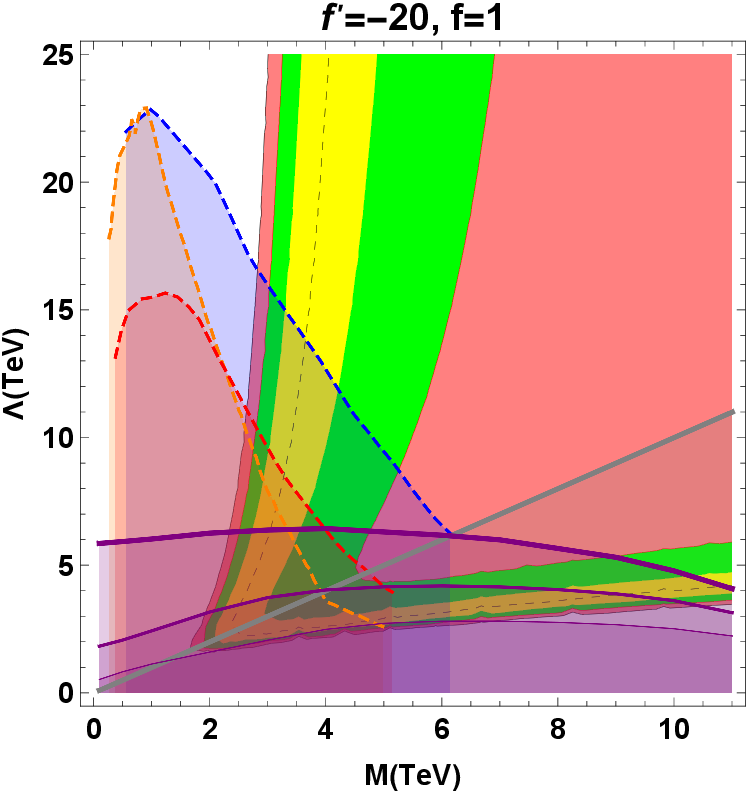  ,scale=0.40,angle=0,clip=}
\psfig{file=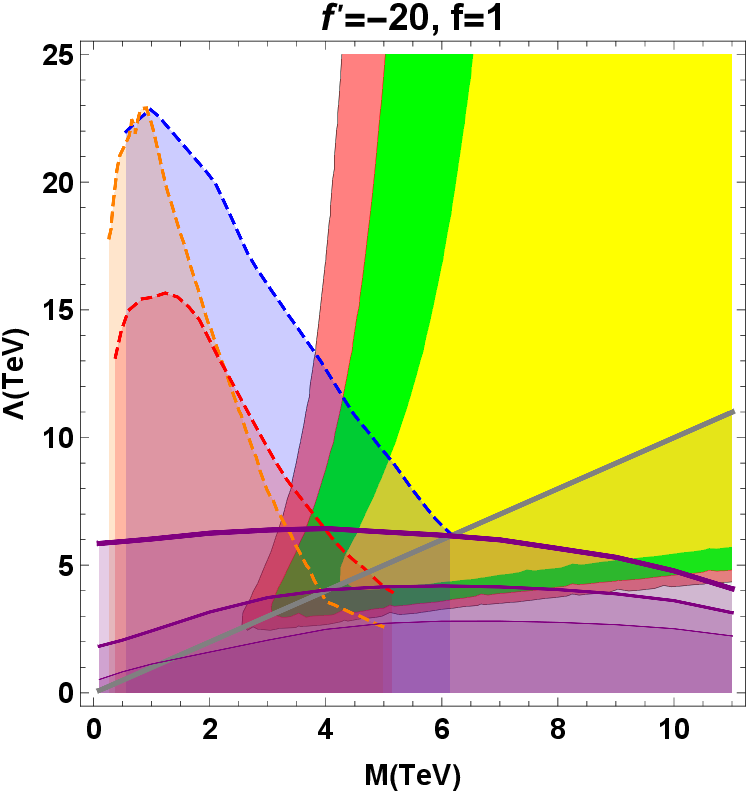 ,scale=0.40,angle=0,clip=}
\end{center}
\caption{$\Delta a_{\mu}^{\rm WA}$ (left plot), $\Delta a_{\mu}^{\rm IB}$ (middle plot ) and $\Delta a_{\mu}^{\rm BMW}$ (right plot) in $(M, \Lambda)$ plane for the case of excited lepton doublet with upper row (lower row) correspond to $f^{\prime}=-10, f=1$ ($f^{\prime}=-20, f=1$). Other lines and color coding remain same as in previous figures.}
\label{fig:DegenDoub-fp1020-f1}
\end{figure}

In \reffi{fig:DegenDoub-fp1020-f1}, we present our findings for $\Delta a_{\mu}^{\rm WA,~IB,~BMW}$ plotted in the $(M, \Lambda)$ plane for the excited lepton doublet, considering different combinations of $f^{\prime}$ and $f$. The plots in the upper row correspond to $f^{\prime}=-10, f=1$ and the plots in the lower row show $\Delta a_{\mu}^{\rm WA,~IB,~BMW}$ for $f^{\prime}=-20, f=1$. While the possibility of selecting negative values for $f$ exists, we have checked that it has minimal impact on the results. Conversely, selecting a negative value for $f^{\prime}$ leads to drastic changes as can be seen in the \reffi{fig:DegenDoub-fp1020-f1}. In this case, it is possible to obtain the required value of $\Delta a_{\mu}^{\rm WA}$ for $f=1$ within the allowed range of parameter space for the doublet contributions. Furthermore, relatively small values of $f^{\prime}$ and $f$ can explain the $\Delta a_{\mu}^{\rm WA}$, contrasting with the previous case where higher values were required. The $\Delta a_{\mu}^{\rm IB}$ and $\Delta a_{\mu}^{\rm BMW}$ can be easily explained as was seen previously.    

\subsection{Contributions to $(g-2)_{\mu}$ for the case of non-degenerate fermion masses}
\label{nondegen-masses}

\subsubsection{Doublet Contributions}
\begin{figure}[htb!]
\begin{center}
\psfig{file=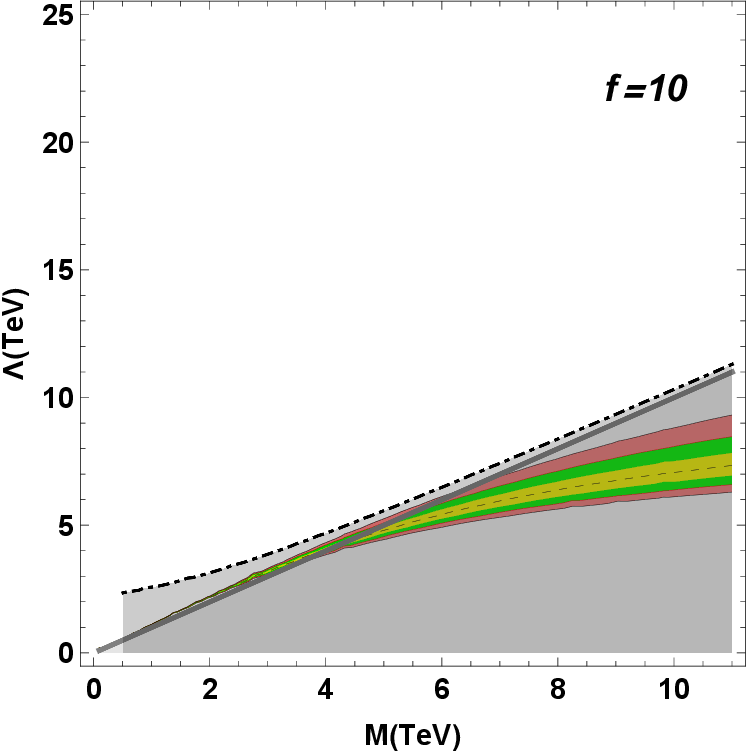  ,scale=0.40,angle=0,clip=}
\psfig{file=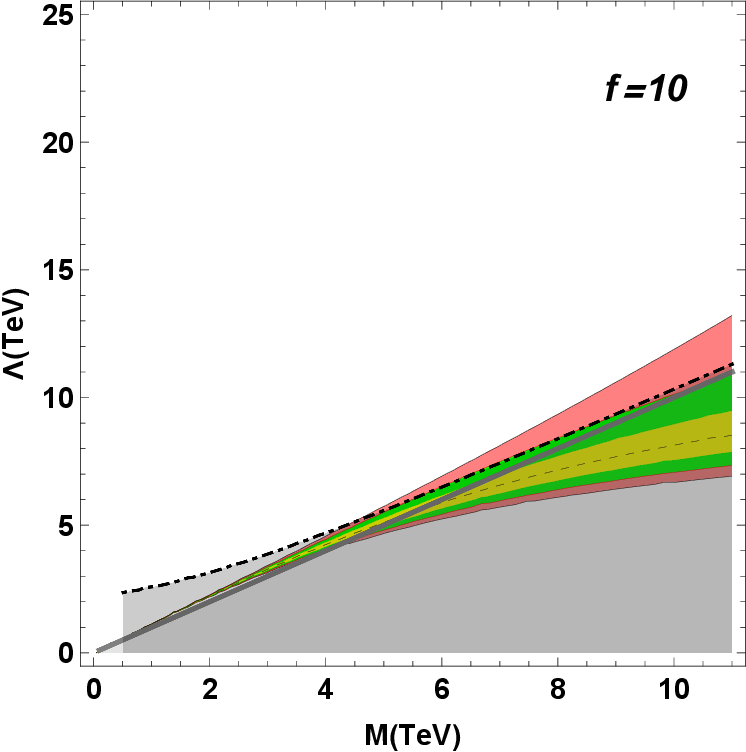  ,scale=0.40,angle=0,clip=}
\psfig{file=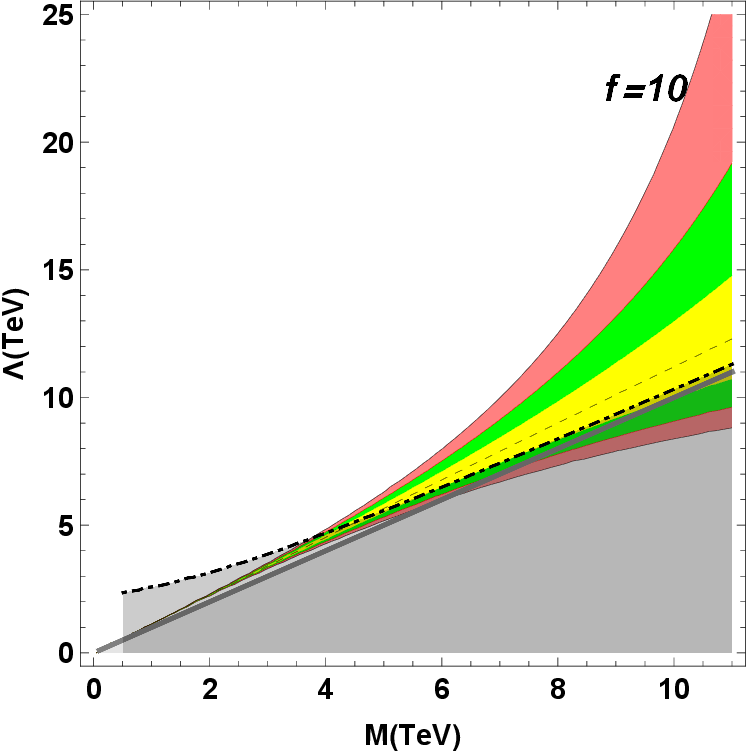  ,scale=0.40,angle=0,clip=}\\
\psfig{file=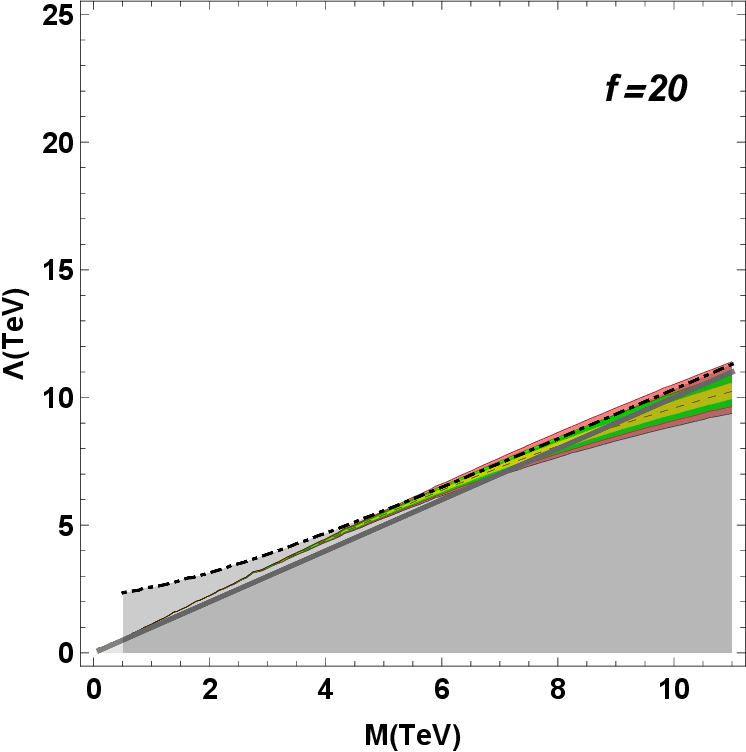 ,scale=0.40,angle=0,clip=}
\psfig{file=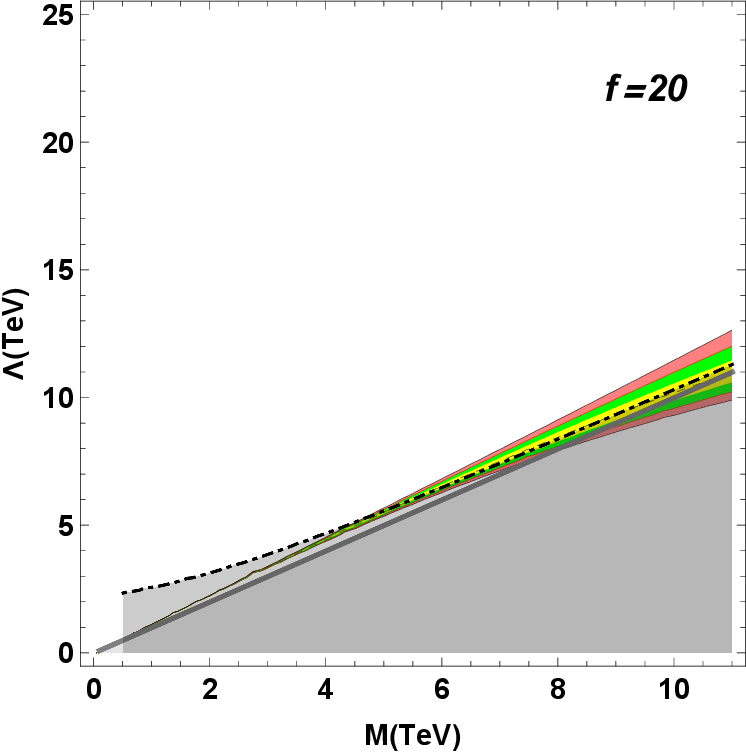 ,scale=0.40,angle=0,clip=}
\psfig{file=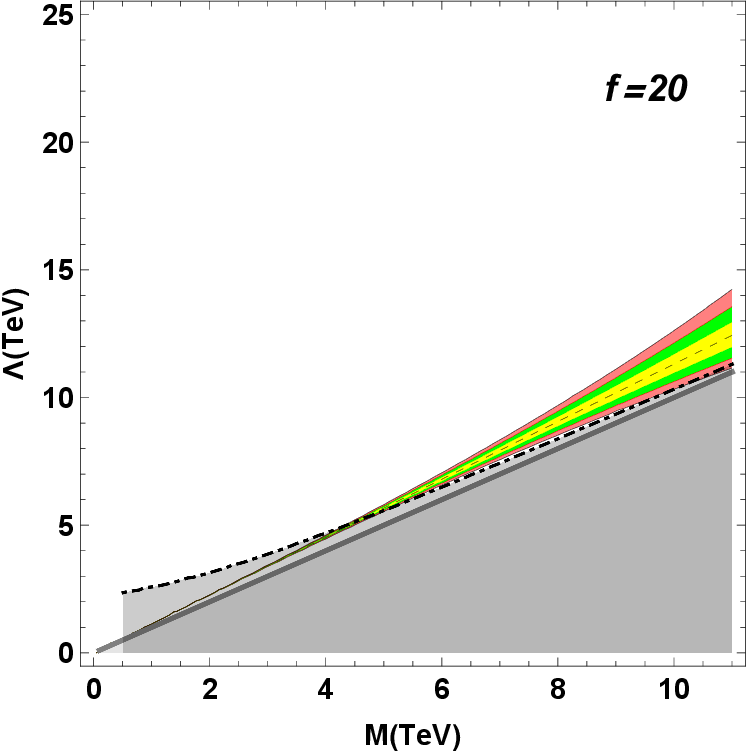 ,scale=0.40,angle=0,clip=}\\
\end{center}
\caption{$\Delta a_{\mu}^{\rm WA}$ (left plot), $\Delta a_{\mu}^{\rm IB}$ (middle plot ) and $\Delta a_{\mu}^{\rm BMW}$ (right plot) in $(M, \Lambda)$ plane for the case of excited lepton doublet with $f=10$ (upper row) and $f=20$ (lower row) including the exclusions from $\Delta \rho$ due to non-degenerate excited fermions with $\delta M=15 \gev$ (dotted-dashed black line). Other lines and color coding remain same as in previous figures.}
\label{fig:NonDegenDoubPlotM15}
\end{figure}

\begin{figure}[htb!]
\begin{center}
\psfig{file=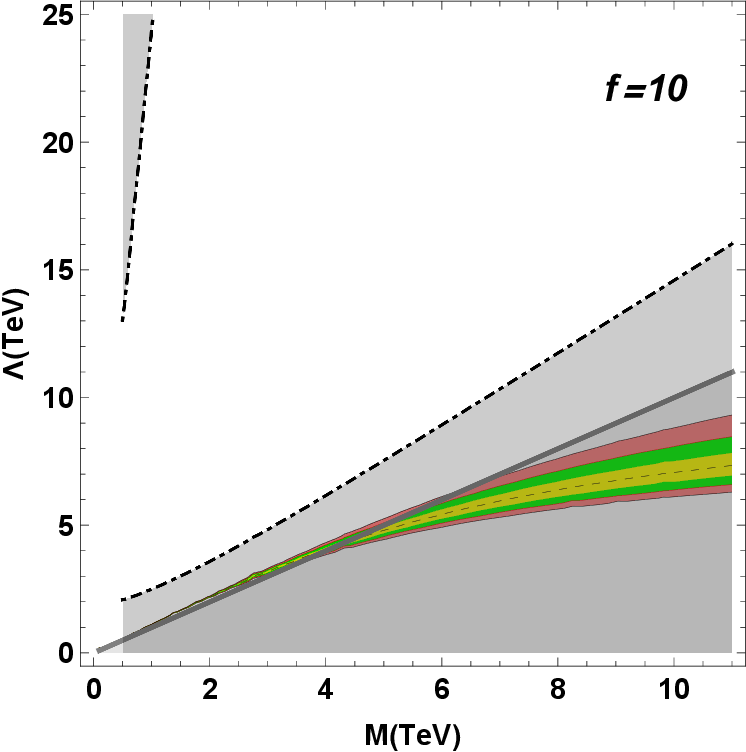  ,scale=0.40,angle=0,clip=}
\psfig{file=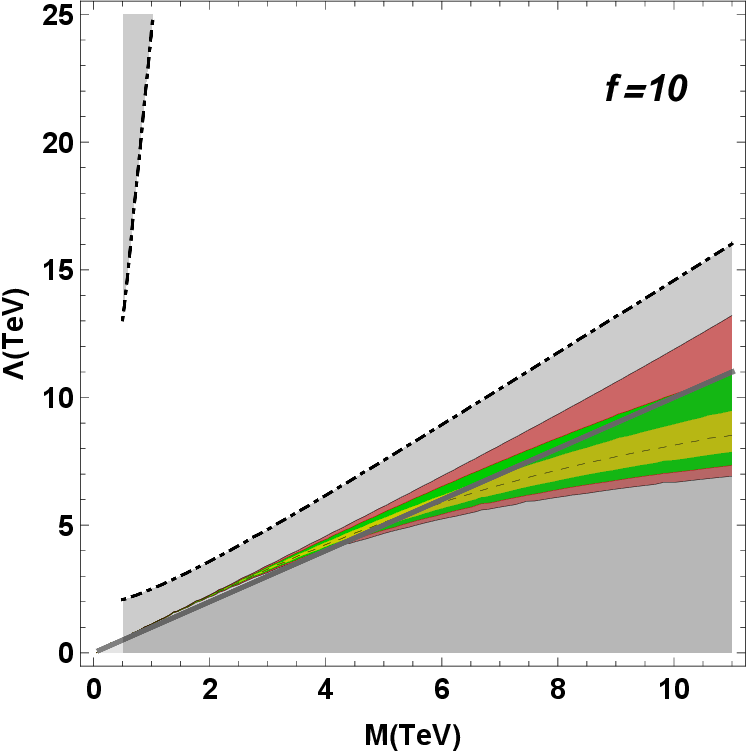  ,scale=0.40,angle=0,clip=}
\psfig{file=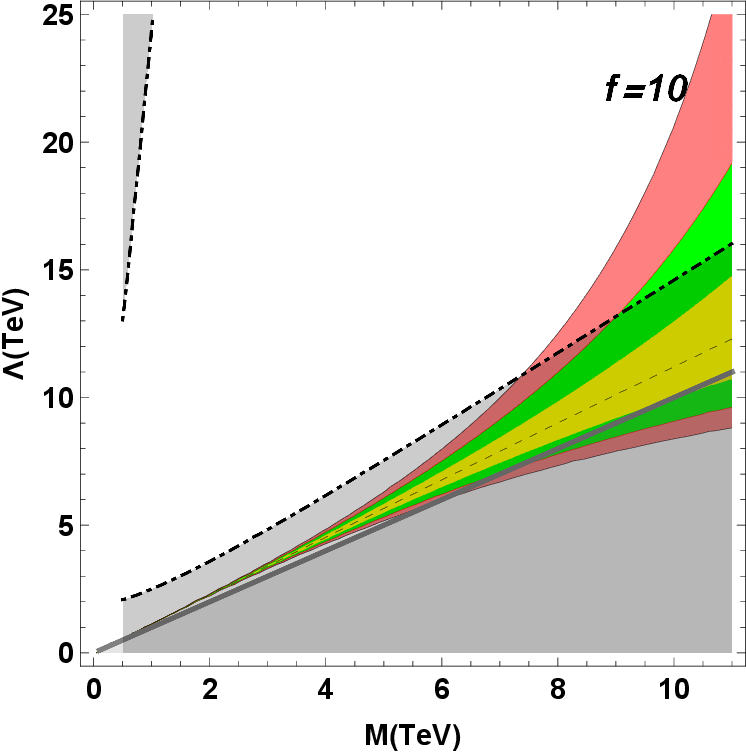  ,scale=0.40,angle=0,clip=}\\
\psfig{file=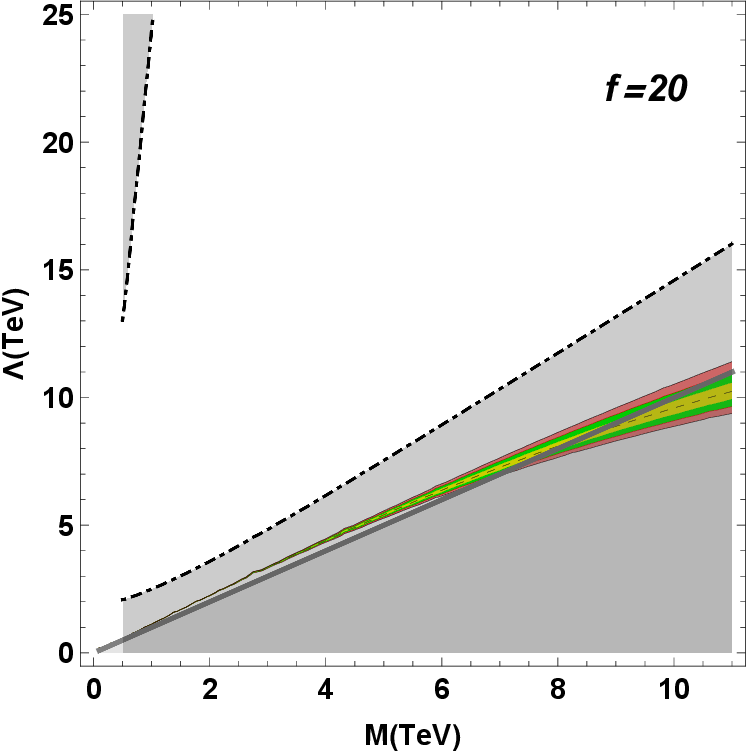 ,scale=0.40,angle=0,clip=}
\psfig{file=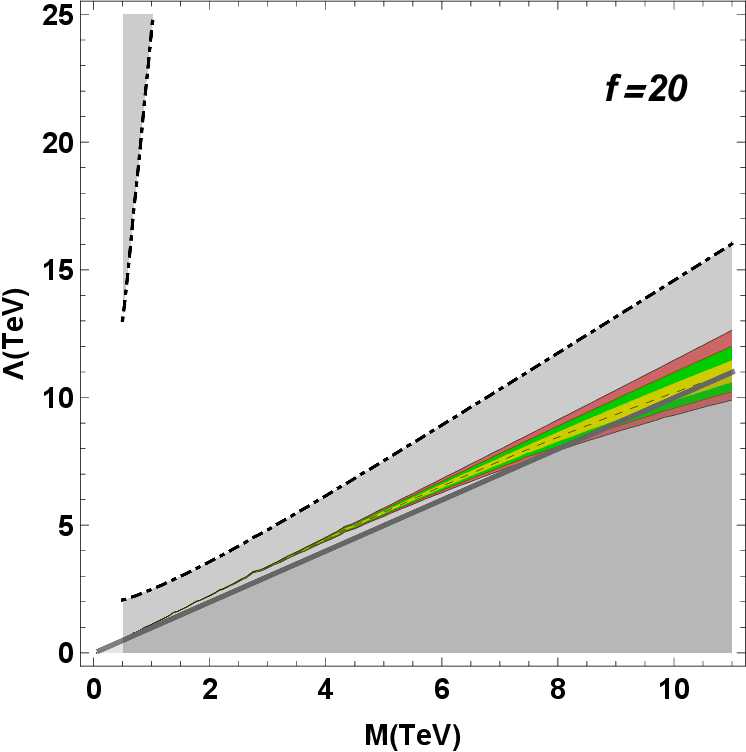 ,scale=0.40,angle=0,clip=}
\psfig{file=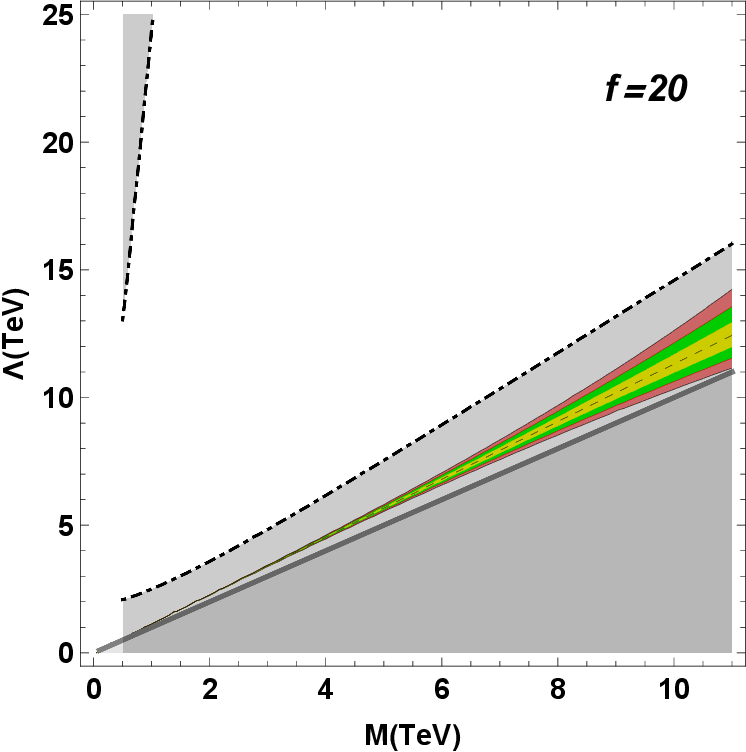 ,scale=0.40,angle=0,clip=} \\
\end{center}
\caption{$\Delta a_{\mu}^{\rm WA}$ (left plot), $\Delta a_{\mu}^{\rm IB}$ (middle plot ) and $\Delta a_{\mu}^{\rm BMW}$ (right plot) in $(M, \Lambda)$ plane for the case of excited lepton doublet with $f=10$ (upper row) and $f=20$ (lower row) including the exclusions from $\Delta \rho$ due to non-degenerate excited fermions with $\delta M=20 \gev$ (dotted-dashed black line). Other lines and color coding remain same as in previous figures.}
\label{fig:NonDegenDoubPlotM20}
\end{figure}

Here, we expand our analysis to investigate the impacts of non-degenerate excited fermions entering via $\Delta \rho$.
For the doublet contributions with $f^{\prime}=f$, we show our results in Figs.~\ref{fig:NonDegenDoubPlotM15}-\ref{fig:NonDegenDoubPlotM20}. As shown in \reffi{fig:NonDegenDoubPlotM15}, the required values of $\Delta a_{\mu}^{\rm WA,~IB,~BMW}$ are difficult to achieve; however, some regions still remain viable if $\delta M<15 \gev$ (dotted-dashed black line), where $\delta M$ denotes the mass difference between the excited fermion species. In contrast, for $\delta M>15 \gev$ (refer to (dotted-dashed black line) in ~\reffi{fig:NonDegenDoubPlotM20}), the $(g-2)_{\mu}$ anomaly cannot be accounted for, as the requisite $\Delta a_{\mu}^{\rm WA,~IB,~BMW}$ value falls within the region excluded by $\Delta \rho$. We also examined cases where $f^{\prime} \neq f$. Nevertheless, the required $(g-2)_\mu$ value remains within the allowed range and is not affected by the $\Delta \rho$ constraints.

\subsubsection{Triplet Contributions}

\begin{figure}[htb!]
\begin{center}
\psfig{file=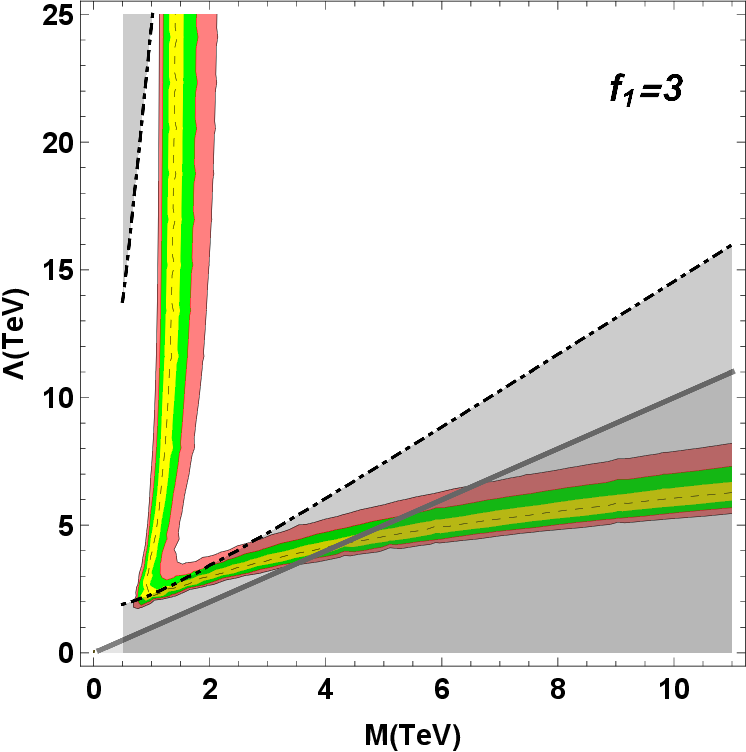  ,scale=0.40,angle=0,clip=}
\psfig{file=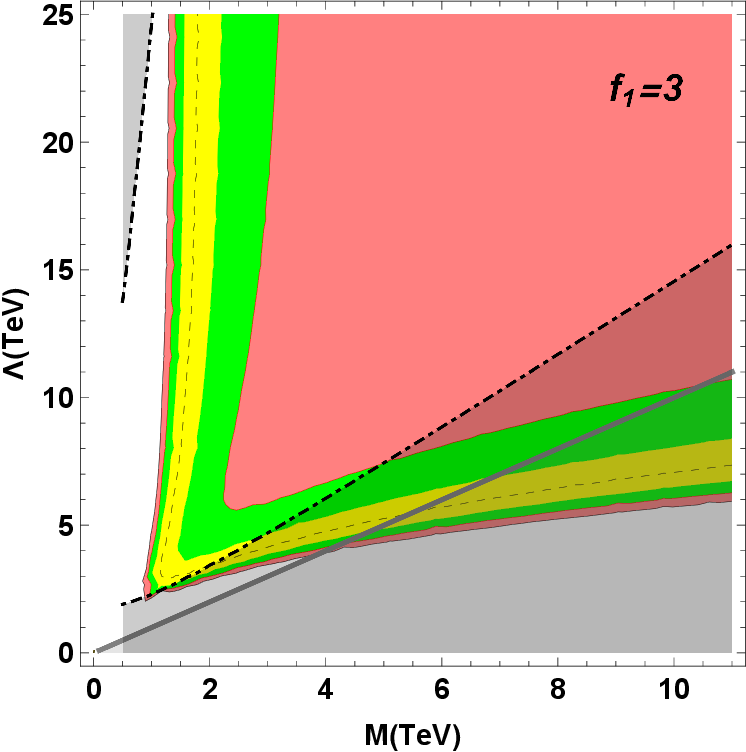  ,scale=0.40,angle=0,clip=}
\psfig{file=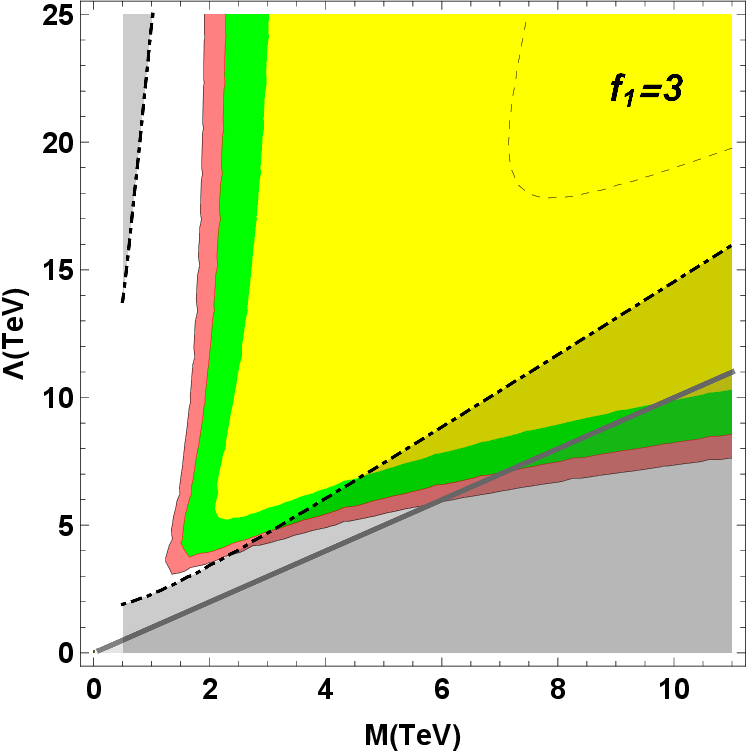  ,scale=0.40,angle=0,clip=}\\
\psfig{file=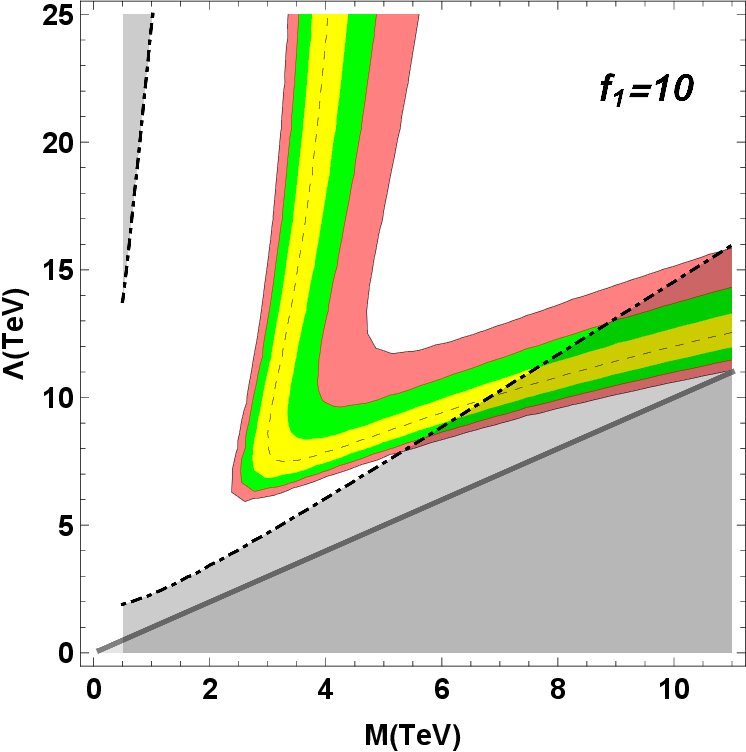 ,scale=0.40,angle=0,clip=}
\psfig{file=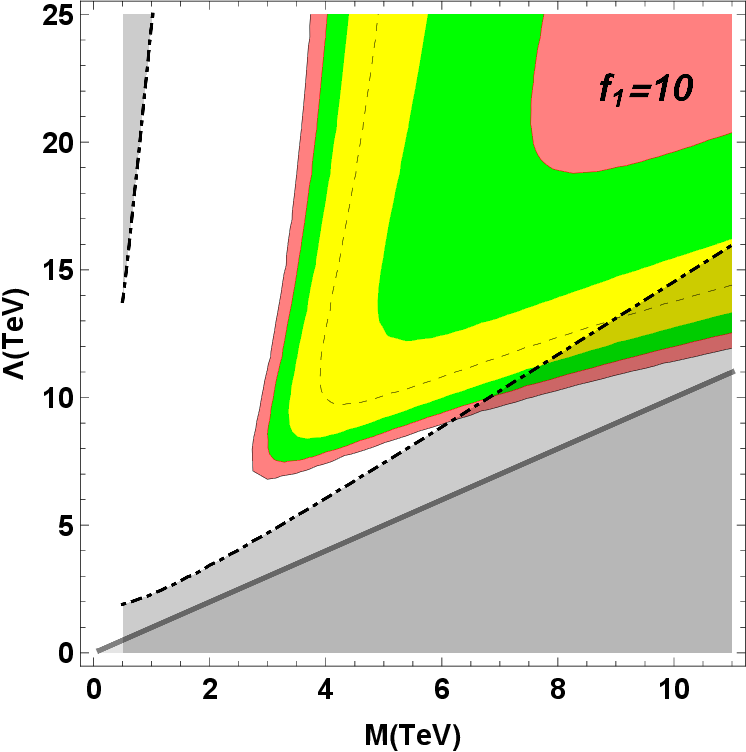 ,scale=0.40,angle=0,clip=}
\psfig{file=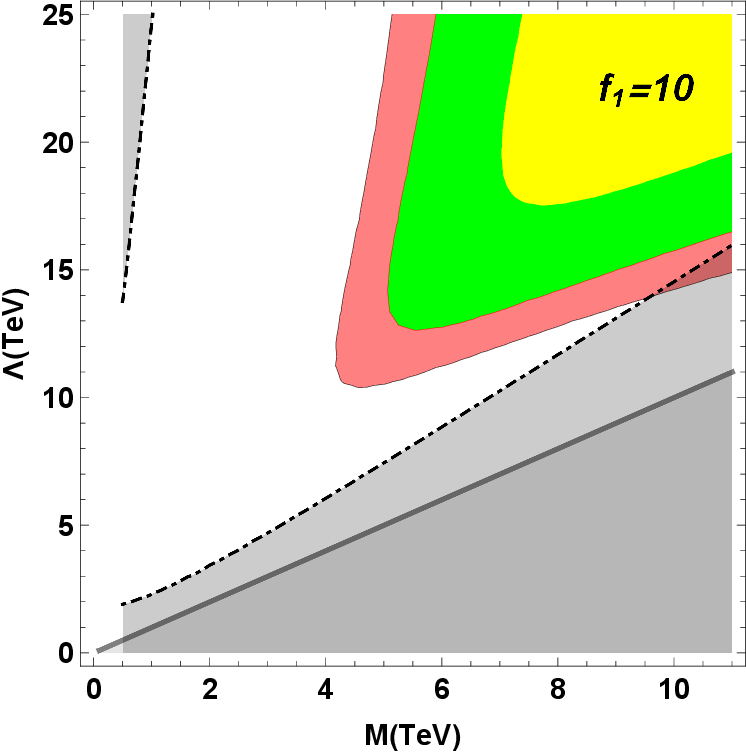 ,scale=0.40,angle=0,clip=}
\end{center}
\caption{$\Delta a_{\mu}^{\rm WA}$ (left plot), $\Delta a_{\mu}^{\rm IB}$ (middle plot ) and $\Delta a_{\mu}^{\rm BMW}$ (right plot) in $(M, \Lambda)$ plane for the case of excited lepton triplets including the exclusions from $\Delta \rho$ due to non-degenerate excited fermions with $\delta M=10 \gev$ (dotted-dashed black line). The upper row corresponds to the value of weight factor $f_1=3$  whereas the lower row correspond to the value $f_1=10$. Other lines and color coding remain same as in previous figures.} 
\label{fig:NonDegenTripPlotM10}
\end{figure}

\begin{figure}[htb!]
\begin{center}
\psfig{file=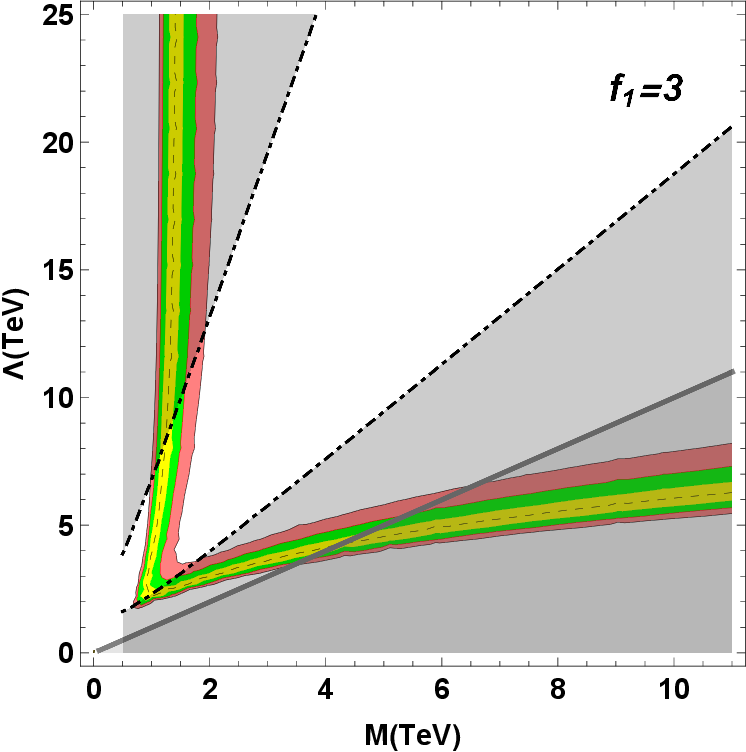  ,scale=0.40,angle=0,clip=}
\psfig{file=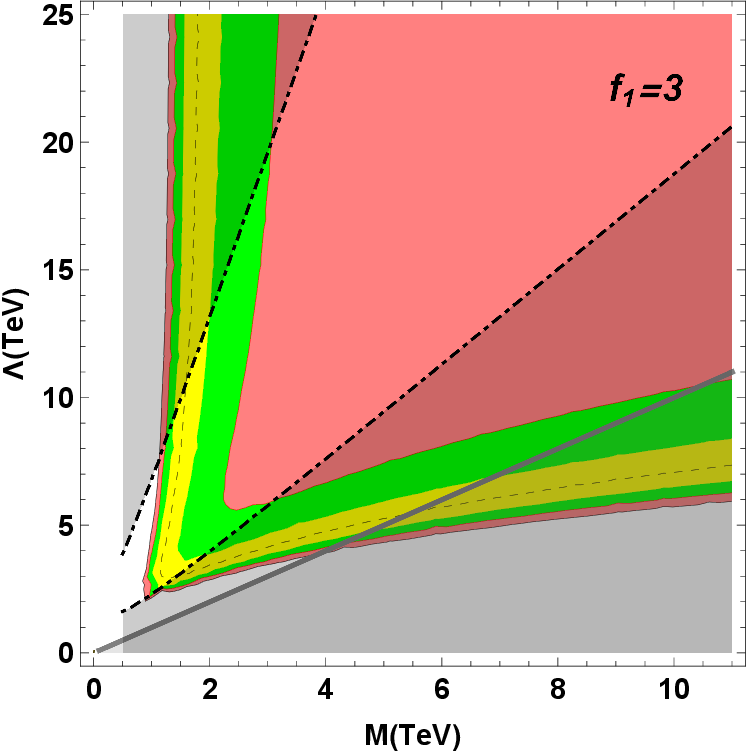  ,scale=0.40,angle=0,clip=}
\psfig{file=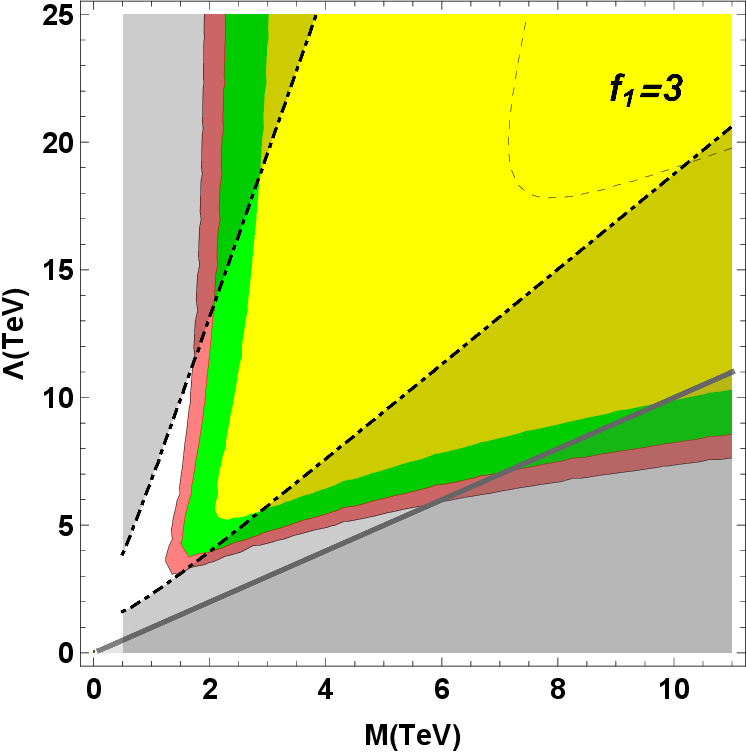  ,scale=0.40,angle=0,clip=}\\
\psfig{file=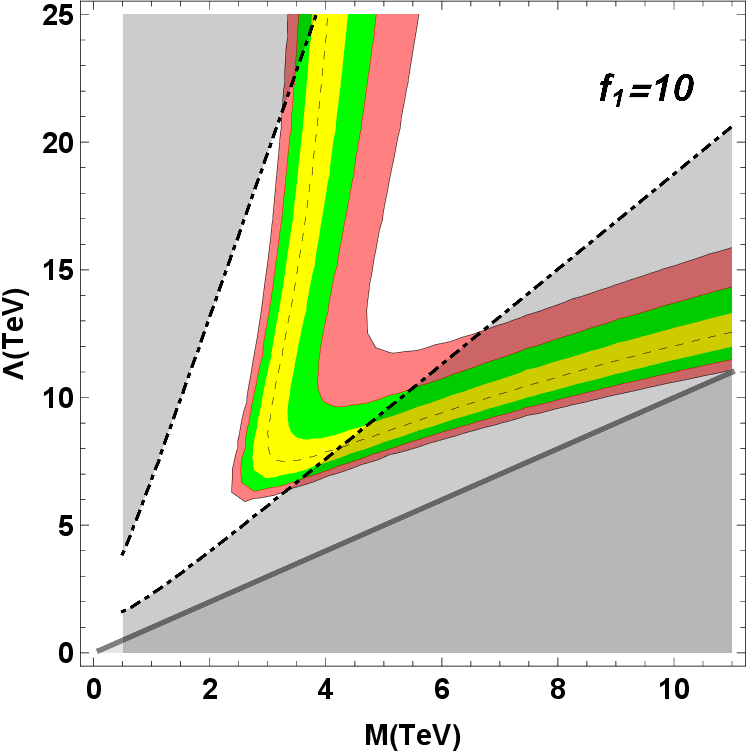 ,scale=0.40,angle=0,clip=}
\psfig{file=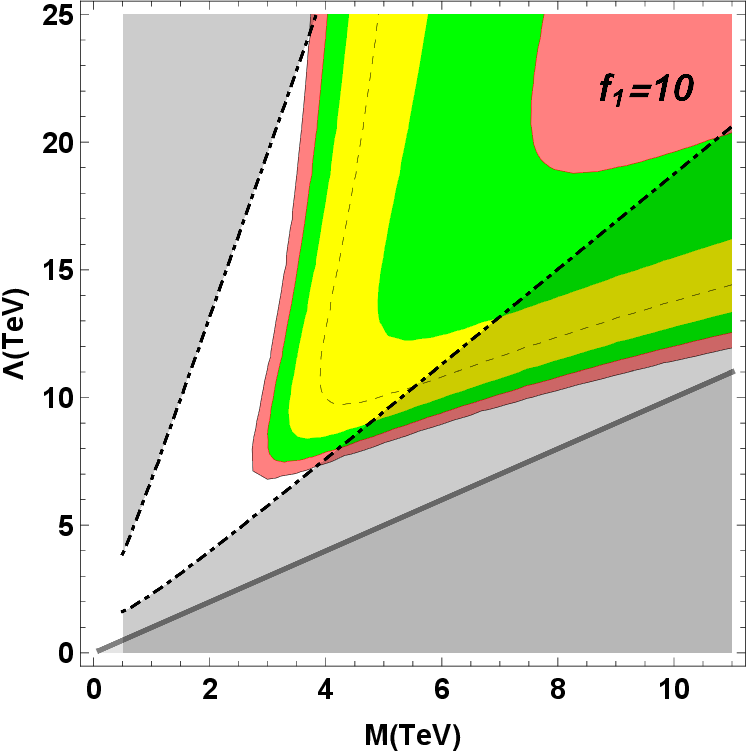 ,scale=0.40,angle=0,clip=}
\psfig{file=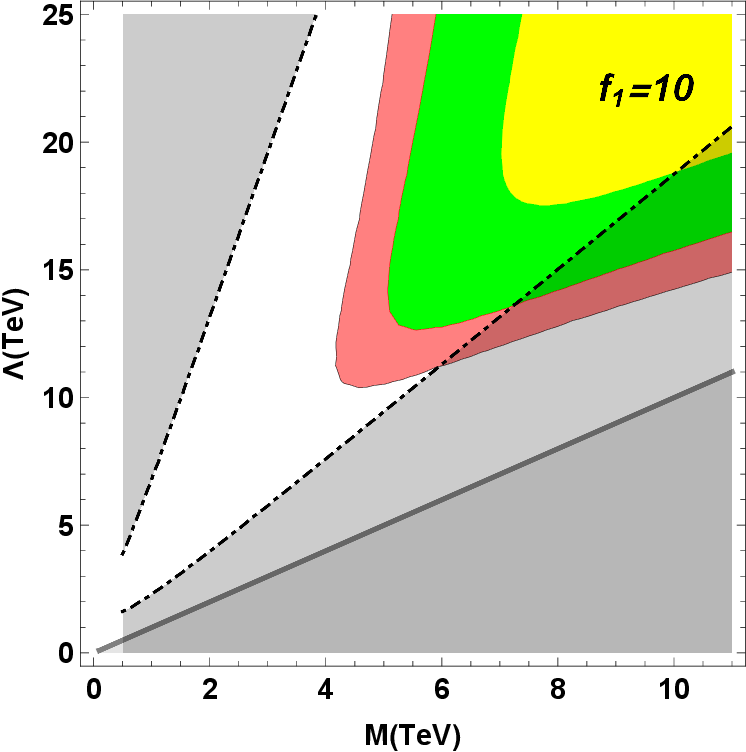 ,scale=0.40,angle=0,clip=}
\end{center}
\caption{$\Delta a_{\mu}^{\rm WA}$ (left plot), $\Delta a_{\mu}^{\rm IB}$ (middle plot ) and $\Delta a_{\mu}^{\rm BMW}$ (right plot) in $(M, \Lambda)$ plane for the case of excited lepton triplets including the exclusions from $\Delta \rho$ due to non-degenerate excited fermions with $\delta M=15 \gev$ (dotted-dashed black line). The upper row corresponds to the value of weight factor $f_1=3$  whereas the lower row correspond to the value $f_1=10$. Other lines and color coding remain same as in previous figures.}
\label{fig:NonDegenTripPlotM15}
\end{figure}

Regarding the contributions from triplets, our findings are presented in Figs.~\ref{fig:NonDegenTripPlotM10}-\ref{fig:NonDegenTripPlotM15}. For $\delta M < 10~\mathrm{GeV}$ (indicated by the dotted-dashed black line), there exists a region in the parameter space where $\Delta a_{\mu}^{\rm WA}$ can be explained. However, for $\delta M > 10~\mathrm{GeV}$, the values of $\Delta a_{\mu}^{\rm WA}$ cannot be accommodated for smaller weight factors, such as $f_1 = 3$. On the other hand, for larger values of $f_1$, such as $f_1 = 10$, certain regions of the parameter space remain viable for $\Delta a_{\mu}^{\rm WA}$.  

In contrast, $\Delta a_{\mu}^{\rm IB}$ and $\Delta a_{\mu}^{\rm BMW}$ retain greater flexibility and can still be achieved within the allowed parameter space, even for larger $\delta M$ values. This highlights the comparatively less stringent dependence of $\Delta a_{\mu}^{\rm IB,~BMW}$ on the weight factor and $\delta M$, providing broader regions that align with the experimental observations. These results underscore the sensitivity of $\Delta a_{\mu}^{\rm WA}$ to the triplet mass splitting $\delta M$ and the weight factor $f_1$, imposing tighter constraints on the model parameters.
 
\clearpage
\section{Conclusions}
\label{sec:conclusions}

The latest measurements of the muon's magnetic moment, $(g-2)_{\mu}$, conducted by the Muon $g-2$ collaboration at Fermilab \cite{Muong-2:2023cdq, Muong-2:2021ojo}, reaffirm that the world average predictions of the Standard Model (SM) for $(g-2)_{\mu}$ are insufficient. However, recent re-analysis of isospin-breaking (IB) corrections to $e^{+}e^{-}$ and $\tau$-decay di-pion observables \cite{Miranda:2024ojv}, along with the latest Budapest-Marseille-Wuppertal (BMW) calculations for hadronic vacuum polarization \cite{Boccaletti:2024guq} and light-by-light scattering contributions \cite{Fodor:2024jyn}, has significantly reduced the discrepancy between SM predictions and experimental results. The tension has now been reduced to $2.7\sigma$ for IB calculations and less than $1\sigma$ for the BMW calculations. This reduction highlights the growing need to explore new physics effects that could bridge the remaining gap between theoretical predictions and experimental observations.

In our previous work, we demonstrated a strong correlation between the prediction for $\Delta \rho$ and the non-degeneracy in mass between excited lepton doublets and triplets. In contrast, the computation of $(g-2)_{\mu}$ is unaffected by mass non-degeneracy. Consequently, we investigated two distinct scenarios: the degenerate scenario, where constraints from $\Delta \rho$ do not apply, and the non-degenerate scenario, where the model's parameters face significant constraints due to $\Delta \rho$.

In this work, we focused on examining the predictions of $(g-2)_{\mu}$ within the excited fermion model, considering higher isospin multiplets with degenerate masses and setting $f^{\prime} = f$. We found that the doublet and triplet contributions to $(g-2)_{\mu}$ are sensitive to the values of the weight factors $f^{\prime}$, $f$, and $f_1$, and show only mild dependence on the weight factor $k$. For the isospin doublet contributions, we observed that the required value of $\Delta a_{\mu}^{\rm WA}$ can only be obtained in a region already excluded by other constraints, unless the weight factor $f$ exceeds $20$. However, $\Delta a_{\mu}^{\rm IB,~BMW}$ can still be explained relatively easily with smaller values of $f$.

The observed deviation in $(g-2)_{\mu}$ can be effectively accounted for by contributions from isospin triplets with smaller values of weight factors. These triplet contributions are particularly significant compared to the doublet contributions, especially when considering equivalent values of the weight factor $f_1$. Moreover, the required $(g-2)_{\mu}$ value can be attained within the allowed ${(M, \Lambda)}$ region for relatively smaller values of $f_1$. As a result, the doublet and triplet contributions to $(g-2)_{\mu}$ impose stringent constraints on the parameter space of the excited fermion model.

We also explored the case where $f^{\prime} \neq f$, with the realization that only $f_1$ affects the triplet contributions. Thus, the results for the triplet contributions remain unchanged in this scenario. However, for the doublets, introducing a negative value for $f^{\prime}$ leads to significant changes. In this case, doublet contributions can also accommodate the $(g-2)_{\mu}$ anomaly within the allowed range of parameter space, with small values for both $f^{\prime}$ and $f$.

Furthermore, we examined the model parameter space for the case of non-degenerate masses, incorporating constraints from experimental searches, unitarity bounds, electroweak precision observable $\Delta \rho$, and $(g-2)_{\mu}$. As reported earlier, a significant portion of the ${(M, \Lambda)}$ plane is excluded by $\Delta \rho$ when the masses are non-degenerate. Consequently, $\Delta a_{\mu}^{\rm WA}$ cannot be explained if the mass difference between the excited fermions, denoted by $\delta M$, exceeds $15\gev$ for excited fermion doublets and $10\gev$ for excited fermion triplets. Nevertheless, $\Delta a_{\mu}^{\rm IB,~BMW}$ can be readily explained even for large values of mass splitting.

\subsection*{Acknowledgments}
The research of M.~E.~G. is supported  by the Spanish MICINN, under grant PID2022-140440NB-C22.
The research of O.~P. is supported by the Istituto Nazionale di Fisica Nucleare, under the grant: ``Exploring New Physics'' (ENP).


\newpage
\pagebreak

\end{document}